\documentclass[twocolumn]{aastex631}

\usepackage{natbib}
\usepackage{subfigure}
\usepackage{mathrsfs}
\usepackage{booktabs}
\usepackage{amsmath}
\usepackage{bm}
\usepackage{morefloats}
\usepackage{multirow}
\usepackage{gensymb}

\newcommand{\xray}{\hbox{\it \rm X-ray\/}}
\newcommand{\xmm}{\hbox{XMM-Newton}}
\newcommand{\chandra}{{Chandra}}
\newcommand{\swift}{{Swift}}
\newcommand{\galex}{{GALEX}}

\newcommand{\aox}{$\alpha_{\rm OX}$}

\graphicspath{{./}{figures/}}

\begin{document}

\title{Strong and Rapid X-ray Variability of the Super-Eddington Accreting Quasar SDSS J081456.10+532533.5}

\author{Jian Huang}
\affiliation{School of Astronomy and Space Science, Nanjing University, Nanjing, Jiangsu 210093, China}
\affiliation{Key Laboratory of Modern Astronomy and Astrophysics (Nanjing University), Ministry of Education, Nanjing 210093, China}

\author{Bin Luo}
\affiliation{School of Astronomy and Space Science, Nanjing University, Nanjing, Jiangsu 210093, China}
\affiliation{Key Laboratory of Modern Astronomy and Astrophysics (Nanjing University), Ministry of Education, Nanjing 210093, China}

\author{W. N. Brandt}
\affiliation{Department of Astronomy \& Astrophysics, 525 Davey Lab,
The Pennsylvania State University, University Park, PA 16802, USA}
\affiliation{Institute for Gravitation and the Cosmos,
The Pennsylvania State University, University Park, PA 16802, USA}
\affiliation{Department of Physics, 104 Davey Lab, The Pennsylvania State University, University Park, PA 16802, USA}

\author{Pu Du}
\affiliation{Key Laboratory for Particle Astrophysics, Institute of High Energy Physics, Chinese Academy of Sciences, 19B Yuquan Road, Beijing 100049, China}

\author{Gordon P. Garmire}
\affiliation{Huntingdon Institute for X-ray Astronomy, LLC, 10677 Franks Road, Huntingdon, PA 16652, USA}

\author{Chen Hu}
\affiliation{Key Laboratory for Particle Astrophysics, Institute of High Energy Physics, Chinese Academy of Sciences, 19B Yuquan Road, Beijing 100049, China}

\author{Hezhen Liu}
\affiliation{School of Astronomy and Space Science, Nanjing University, Nanjing, Jiangsu 210093, China}
\affiliation{Key Laboratory of Modern Astronomy and Astrophysics (Nanjing University), Ministry of Education, Nanjing 210093, China}

\author{Qingling Ni}
\affiliation{Max-Planck-Institut f\"{u}r extraterrestrische Physik (MPE), Gie{\ss}enbachstra{\ss}e 1, D-85748 Garching bei M\"unchen, Germany}

\author{Jian-Min Wang}
\affiliation{Key Laboratory for Particle Astrophysics, Institute of High Energy Physics, Chinese Academy of Sciences, 19B Yuquan Road, Beijing 100049, China}
\affiliation{School of Astronomy and Space Science, University of Chinese Academy of Sciences, 19A Yuquan Road, Beijing 100049, China}
\affiliation{National Astronomical Observatories of China, Chinese Academy of Sciences, 20A Datun Road, Beijing 100020, China}

\begin{abstract}

We report strong and rapid \xray\ variability found from the \hbox{super-Eddington} accreting quasar SDSS J081456.10+532533.5 at $z=0.1197$.
It has a \hbox{black-hole} mass of $2.7\times10^{7}{M_{\odot}}$ and a dimensionless accretion rate of $\approx4$ measured from reverberation-mapping observations.
It showed weak \xray\ emission in the 2021 February \chandra\ observation, with the \hbox{2 keV} flux density being $9.6^{+11.6}_{-4.6}$ times lower compared to an archival \swift\ observation.
The \hbox{2 keV} flux density is also $11.7^{+9.6}_{-6.3}$ times weaker compared to the expectation from its optical/UV emission.
In a \hbox{follow-up} \xmm\ observation 32 days later, the \hbox{2 keV} flux density increased by a factor of $5.3^{+6.4}_{-2.4}$, and the spectra are best described by a power law modified with \hbox{partial-covering} absorption; the \hbox{absorption-corrected} intrinsic continuum is at a nominal flux level.
Nearly simultaneous optical spectra reveal no variability, and there is only mild \hbox{long-term} optical/infrared variability from archival data (with a maximum variability amplitude of $\approx50\%$).
We interpret the \xray\ variability with an obscuration scenario, where the intrinsic \xray\ continuum does not vary but the absorber has variable column density and covering factor along the line of sight.
The absorber is likely the \hbox{small-scale} clumpy accretion wind that has been proposed to be responsible for similar \xray\ variability in other \hbox{super-Eddington} accreting quasars.

\end{abstract}

\section{Introduction} \label{sec:intro}
Active galactic nuclei (AGNs) are powered by accretion onto supermassive black holes (SMBHs).
The optical/UV radiation of an AGN is likely dominated by the accretion-disk thermal emission (e.g., \citealt{Shakura1973}), while the \xray\ emission is considered to be largely produced in the accretion-disk corona via inverse Compton scattering of the optical/UV radiation (e.g., \citealt{Sunyaev1980,Haardt1993}). 
Observations have revealed significant correlations between the \xray\ and optical/UV radiation in radio-quiet type 1 AGNs, such as the negative correlation between the \hbox{\xray-to-optical} \hbox{power-law} slope parameter ($\alpha_{\rm OX}$\footnote{$\alpha_{\rm OX}$ is defined as $\alpha_{\rm OX}=-0.3838~{\rm log}(f_{2500~\textup{\AA}}/f_{\rm 2~keV})$, where $f_{2500~\textup{\AA}}$ and $f_{\rm 2~keV}$ are the flux densities at \hbox{2500 \AA} and \hbox{2 keV}, respectively.}) and 2500 \AA\ monochromatic luminosity ($L_{2500~\textup{\AA}}$; e.g., \citealt{Steffen2006,Just2007,Pu2020}) and the non-linear correlation between the \xray\ and optical/UV monochromatic luminosities ($L_{\rm 2~keV}\textrm{--}L_{2500~\textup{\AA}}$; e.g., \citealt{Risaliti&Lusso2019}).
These correlations indicate a physical connection between the accretion disk and the corona. 
Deviations from these correlations have been found in various samples of type 1 AGNs, usually showing weaker \xray\ emission than that expected from the correlations (e.g., \citealt{Luo2015,Nardini2019,Pu2020,Timlin2020, Laurenti2022,Ni2022}; but see \citealt{Fu2022} for a small sample of quasars with excess \xray\ emission).
An observed deficit of \xray\ emission often indicates \xray\ obscuration
from dust-free absorbers (which do not greatly attenuate optical/UV radiation), but there have also been proposals of intrinsic \xray\ weakness for some objects (e.g., \citealt{Leighly2007b,Leighly2007a,Luo2014}; but see \citealt{Wang2022} for an obscuration scenario).

X-ray variability is also a characteristic feature of AGNs, which generally has larger amplitudes and shorter timescales than AGN optical/UV variability.
The typical \xray\ flux variability amplitudes for radio-quiet type 1 AGNs ranges from $\approx20\%$ to 50$\%$, which is generally ascribed to instability/fluctuation of the accretion disk and the corona
\citep[e.g.,][]{McHardy2006,MacLeod2010,Yang2016,Zheng2017}.
Strong \xray\ variability (flux variability amplitude $>2$) events are rare
(e.g., \citealt{Mushotzky1993,Gibson2012,Middei2017,Timlin2020}), and they usually require significant changes of the disk/corona properties or interpretations external to the disk/corona (e.g., \xray\ obscuration).
For example, some ``\hbox{changing-look}" AGNs show strong X-ray variability on timescales of years, which might be attributed to changes of accretion rates \citep[e.g.,][]{LaMassa2015,Parker2016,Mathur2018,Oknyansky2019}.
Another example is the strong and sometimes rapid (down to minutes) \xray\ variability observed from some \hbox{narrow-line} Seyfert 1 galaxies (NLS1s), and one interpretation is that they are affected by variable \xray\ obscuration from clumpy \hbox{accretion-disk} winds (e.g., \citealt{Reeves2019,Boller2021,Parker2021,Jin2022}).
NLS1s are considered to have high or even \hbox{super-Eddington} accretion rates,
and powerful disk winds launched via radiation pressure are generally 
expected in these systems (e.g., \citealt{Jiang2014,Sadowski2014,Jiang2019}).
Strong \xray\ variability events have also been observed in a few 
typical type 1 AGNs (e.g., NGC 3227, \citealt{Mao2022,Wangyijun2022}; NGC 5548, \citealt{Kaastra2014,Cappi2016,Mehdipour2022}), and they were also attributed to \hbox{dust-free} \xray\ obscuration.

Recently, an increasing number of type 1 quasars have been discovered to 
exhibit strong and sometimes rapid (down to hours) \xray\ variability
(e.g., PHL 1092, \citealt{Miniutti2012}; SDSS J075101.42+291419.1, \citealt{Liu2019}; SDSS J153913.47+395423.4, \citealt{Ni2020}; and SDSS J135058.12+261855.2, \citealt{Liu2022}), similar to that in NLS1s.
There is no simultaneous optical/UV continuum or \hbox{emission-line} variability, indicating no significant changes in the accretion rate.
These quasars varied between \xray\ nominal-strength states and multiple \xray\ weak states in the $\alpha_{\rm OX}$ versus $L_{2500~\textup{\AA}}$ plane.
They are also considered to have high or even \hbox{super-Eddington} accretion rates.
These features suggest that the strong \xray\ variability is likely also related to \hbox{dust-free} obscuration from \hbox{accretion-disk} winds.

In this paper, we report another low-redshift (\hbox{$z=0.1197$}), type 1 quasar that shows strong and rapid \xray\ variability, SDSS J081456.10+532533.5 (hereafter SDSS \hbox{J$0814+5325$}).
It has a SMBH mass of $2.7\times10^{7}{M_{\odot}}$ measured from reverberation-mapping observations \citep{Du2015}.
Its high dimensionless accretion rate ($\dot{\rm \mathscr{M}}\approx4$) \footnote{$\dot{\rm \mathscr{M}}=\dot{M}/({L_{\rm Edd}c^{2})}=20.1{(L_{5100~\textup{\AA}}/10^{44}/{\rm cos}~i)}^{3/2}{(M_{\rm BH}/{10}^{7}M_{\odot})}^{-2}$ \citep{Shakura1973}, where $\dot{M}$ is the \hbox{mass-accretion} rate, $L_{\rm Edd}$ is the Eddington luminosity, $L_{5100~\textup{\AA}}$ is the AGN luminosity at 5100 \AA, $i$ is the inclination angle, and $M_{\rm BH}$ is the SMBH mass.} satisfies the criterion of $\dot{\rm \mathscr{M}}>3$ commonly used for identifying \hbox{super-Eddington} accreting AGNs \citep[e.g.,][]{Du2015,Du2016}.
Its unusual \xray\ variability was revealed by our new \chandra\ and \xmm\ observations.
The \hbox{5100 \AA} luminosity (${\rm log}~[L_{5100~{\textup{\AA}}}/{\rm erg}~{\rm s}^{-1}]\approx44.0$) of SDSS \hbox{J$0814+5325$} is relatively low among quasars, bridging the gap between NLS1s and quasars showing strong \xray\ variability (e.g., Table 4 of \citealt{Liu2019}).
We organize our study as follows.
In Section~\ref{sec:2}, we describe the \xray\ and multiwavelength observations of SDSS \hbox{J$0814+5325$}.
In Section~\ref{sec:3}, we present \xray\ spectral properties, \xray\
variability, and multiwavelength properties.
In Section~\ref{sec:4}, we discuss the strong and rapid \xray\ variability in the obscuration scenario.
We summarize in Section~\ref{sec:5}.
Throughout this paper, we use a cosmology of \hbox{$H_{0}=67.4~{\rm km}~{\rm s}^{-1}~{\rm {Mpc}}^{-1}$}, $\Omega_{\rm M} = 0.315$, and $\Omega_{\Lambda} = 0.685$ \citep{Planck2020}.

\begin{deluxetable*}{lcccc}
\tablewidth{0pt}
\tablecaption{\xray\ Observation Log}
\tablehead{
\colhead{Observatory}  &
\colhead{Observation}  &
\colhead{Observation} &
\colhead{Exposure Time}   &
\colhead{Comment}  \\
\colhead{ }   &
\colhead{ID}   &
\colhead{Start Date}   &
\colhead{(ks)}   &
\colhead{} \\
\colhead{(1)}   &
\colhead{(2)}   &
\colhead{(3)}   &
\colhead{(4)}   &
\colhead{(5)}
}
\startdata
\multirow{3}{*}{Swift}      & 00039566001  & 2010 Jan 17 & 1.9  & UVW2(S) \\
    & 00039566002  & 2011 Mar 06 & 1.0  & UVM2(S) \\
    & 00039566003  & 2011 Aug 24 & 2.2  & UVW2(S)  \\
{Chandra}      & 24727  & 2021 Feb 25 & 1.3  &  \\
{XMM-Newton} & 0872392101   & 2021 Apr 03 & 20.5 (pn) & UVW2 UVM2 UVW1 U \\
\enddata
\tablecomments{Column (1): name of the \xray\ observatory. Column (2): observation ID. Column (3): observation start date. Column (4): cleaned exposure time. Column (5): the simultaneous UV observations available.}
\label{tbl:obslog}
\end{deluxetable*}

\section{X-ray and Multiwavelength Observations} \label{sec:2}

\subsection{Chandra Observation} \label{sec:2.1}

SDSS \hbox{J$0814+5325$} was observed by \chandra\ \hbox{ACIS-S} on 2021 February 25 as part of our Chandra snapshot program searching for strong \xray\ variability among \hbox{reverberation-mapped} \hbox{super-Eddington} accreting AGNs.
The basic information for this \chandra\ observation is given in Table~\ref{tbl:obslog}.
The exposure time is \hbox{1.3 ks}.
The \xray\ properties of the other seven targets in the snapshot program are presented in the Appendix.

We processed the data using the Chandra Interactive Analysis of Observation (CIAO; v4.12) tools.
The {\sc chandra\_repro} script was first used to generate a new level 2 event file.
Then we ran the {\sc deflare} script and used an iterative $3\sigma$ clipping algorithm to filter potential background flares; no flare was identified and the cleaned exposure time is still \hbox{1.3 ks}.
We used the {\sc dmcopy} tool to create \xray\ images in the \hbox{0.3--2}, \hbox{2--8} and \hbox{0.3--8} keV bands.
We used the {\sc wavdetect} tool \citep{Freeman2002} with a \hbox{false-positive} probability threshold of $10^{-6}$ and scale sizes of 1, 1.414, 2, 2.828, 4, 5.656, 8 pixels to search for \xray\ sources in the three images.
SDSS \hbox{J$0814+5325$} was detected in all three bands.
In the \hbox{0.3--8} keV band, its \xray\ positional offset to its optical position is 0.7\arcsec.
We used the \hbox{{\sc specextract}} tool to extract source and background spectra.
A circular source region with a radius of 2\arcsec\ centered on its \xray\ position was chosen to extract the source spectrum, and an annulus with radii of 6\arcsec\ and 10\arcsec\ centered on the \xray\ position was used as the background region.
We verified that the background region does not contain any \xray\ sources.
The spectrum has $\approx12$ net counts in the \hbox{0.3--8 keV} band, and it was grouped with at least one count per bin for spectral fitting.

\begin{deluxetable*}{llrcccccrrc}
\tablewidth{0pt}
\tablecaption{Optical/UV Measurements}
\tablehead{
\colhead{Observatory}  &
\colhead{Observation}  &
\colhead{$f_{\rm UVW2}$} &
\colhead{$f_{\rm UVM2}$} &
\colhead{$f_{\rm UVW1}$} &
\colhead{$f_{\rm U}$} &\\
\colhead{ }   &
\colhead{Start Date}   &
\colhead{ }   &
\colhead{ }   &
\colhead{ }   &
\colhead{ }   &
\colhead{ }  \\
\colhead{(1)}   &
\colhead{(2)}   &
\colhead{(3)}   &
\colhead{(4)}   &
\colhead{(5)}   &
\colhead{(6)}  
}
\startdata
\multirow{3}{*}{Swift} & 2010 Jan 17  & $1.39^{+0.04}_{-0.04}$ & ... & ... & ... \\
 & 2011 Mar 06  & ... & $1.77^{+0.07}_{-0.07}$ & ... & ...  \\
 & 2011 Aug 24  & $2.18^{+0.05}_{-0.05}$ & ... & ... & ...  \\
{ XMM-Newton} & 2021 Apr 03  & $1.99^{+0.09}_{-0.09}$ & $2.19^{+0.08}_{-0.08}$ & $3.63^{+0.03}_{-0.03}$ & $4.42^{+0.04}_{-0.04}$  \\
\enddata
\tablecomments{
Column (1): \xray\ observatory.
Column (2): observation start date.
Columns (3)--(6): \hbox{Galactic-extinction-corrected} \swift\ UVOT or \xmm\ OM \hbox{flux-density} measurements in units of $10^{-27}~{\rm erg}~{\rm cm}^{-2}~{\rm s}^{-1}~{\rm Hz}^{-1}$.
}
\label{tbl:x_uv_o}
\end{deluxetable*}

\subsection{XMM-Newton Observation} \label{sec:2.2}

Due to the strong \xray\ variability revealed by the \chandra\ observation compared to the archival \swift\ observations (see Section~\ref{sec:3.3} below), an \xmm\ Target of Opportunity (TOO) observation was triggered about a month later to obtain a deeper \xray\ exposure ($\approx28~{\rm ks}$ total exposure).
The basic information of the \xmm\ observation is presented in Table~\ref{tbl:obslog}.

The Science Analysis System (SAS; v19.0.0) was used for the \xmm\ data reduction.
We followed the standard procedure in the SAS Data Analysis Threads\footnote{\url{https://www.cosmos.esa.int/web/xmm-newton/sas-threads}.} to process the EPIC pn \citep{Struder2001} and EPIC MOS (MOS1 and MOS2; \citealt{Turner2001}) data.
We obtained calibrated and concatenated event lists using the task {\sc epproc} for the pn detector and {\sc emproc} for the MOS detectors.
\hbox{Count-rate} thresholds of \hbox{0.4 ${\rm cts}~{\rm s}^{-1}$} (pn) and \hbox{0.35 ${\rm cts}~{\rm s}^{-1}$} (MOS) were used to filter background flares, and the task {\sc tabgtigen} was used to create \hbox{good-time-interval} files.
The cleaned exposure times for the pn, MOS1, and MOS2 data are \hbox{20.5 ks}, \hbox{26.7 ks}, and \hbox{26.7 ks}, respectively.
We then created cleaned event files using the {\sc evselect} task.
For each detector, source and background spectra were extracted using the task {\sc evselect} with a \hbox{30\arcsec-radius} circular source region and a \hbox{40\arcsec-radius} circular \hbox{source-free} background region on the same CCD chip.
The pn, MOS1, and MOS2 spectra have $7306$, $2334$, and $2301$ net counts in the \hbox{0.3--10 keV} band, respectively.
We grouped the source spectra with at least 25 counts per bin for spectral fitting.

Four Optical Monitor (OM; \citealt{Mason2001}) optical/UV filters were used in this observation, including UVW2, UVM2, UVW1, and U with effective wavelengths of 2120~\AA, 2310~\AA, 2910~\AA, and 3440~\AA, respectively.
We used the pipeline task {\sc omichain} to process the OM data. 
Source flux and magnitude for each filter were extracted from the generated SWSRLI files.
We adopted the mean magnitudes and fluxes of all the exposure segments for each filter.
We de-reddened the UV fluxes adopting the \cite{Fitzpatrick2019} Milky Way \hbox{$R_V$-dependent} extinction model with $R_V=3.1$ and Galactic extinction \hbox{$E(B-V)=0.032$} \citep{Schlegel1998}.
The OM measurements are listed in Table~\ref{tbl:x_uv_o}.
We then derived the \hbox{2500~\AA} flux density, $f_{2500~\textup{\AA}}=3.33\times10^{-27}~{\rm erg}~{\rm cm}^{-2}~{\rm s}^{-1}~{\rm Hz}^{-1}$, that is simultaneous to the EPIC \xray\ observation via an interpolation of the UVM2 and UVW1 flux densities.
This $f_{2500~\textup{\AA}}$ value was corrected for intrinsic extinction in Section~\ref{sec:3.2} below and it was then used for the \aox\ computations.

\subsection{Archival Swift Observations} \label{sec:2.3}

There are three archival \swift\ \xray\ Telescope (XRT; \citealt{Burrows2005}) observations of SDSS \hbox{J$0814+5325$}, listed in Table~\ref{tbl:obslog}.
The Photon Counting (PC) mode was used in these observations.
The task {\sc xrtpipeline}, which is included in the HEASOFT 6.27.2 package, was used for data reduction of the XRT observations.
SDSS \hbox{J$0814+5325$} was significantly detected in the \hbox{0.3--8 keV} band in all these \swift\ observations.
For each observation, we used the task {\sc xselect} to extract a source spectrum from a circular region with a radius of 47\arcsec\ centered on the optical position.
A background spectrum was extracted from a nearby \hbox{source-free} circular region with a radius of 100\arcsec.
We used the task {\sc xrtmkarf} to generate an ancillary response function file and we obtained the standard photon redistribution matrix file from the HEASARC's calibration database (CALDB).
The three XRT spectra have $\approx31\textrm{--}74$ net counts in the \hbox{0.3--8 keV} band, and they were grouped with at least one count per bin.

The XRT observations have simultaneous Swift \hbox{UV-Optical} Telescope (UVOT; \citealt{Roming2005}) observations.
Only one filter was used in each UVOT observation (Table~\ref{tbl:obslog}).
Two of the observations have the UVW2(S) filter with an effective wavelength of \hbox{2030~\AA} and one has the UVM2(S) filter with an effective wavelength of \hbox{2231~\AA} \citep{Poole2008}.
For each UVOT observation, we used the task {\sc uvotimsum} to stack the \hbox{aspect-corrected} images.
The task {\sc uvotdetect} was used to determine the UV position of SDSS \hbox{J$0814+5325$} and to search for other UV sources in the stacked image.
A circular source region with a radius of 5\arcsec\ centered on the UV position was used to extract source counts from the stacked image.
A nearby \hbox{source-free} region with a radius of 20\arcsec\ was chosen to extract background counts.
We then used the task \hbox{{\sc uvotsource}} to perform aperture photometry.
The derived UV flux density was corrected for the Galactic extinction.
The UVOT measurements are listed in Table~\ref{tbl:x_uv_o}.

\begin{figure}
\centering
\includegraphics[scale=0.33]{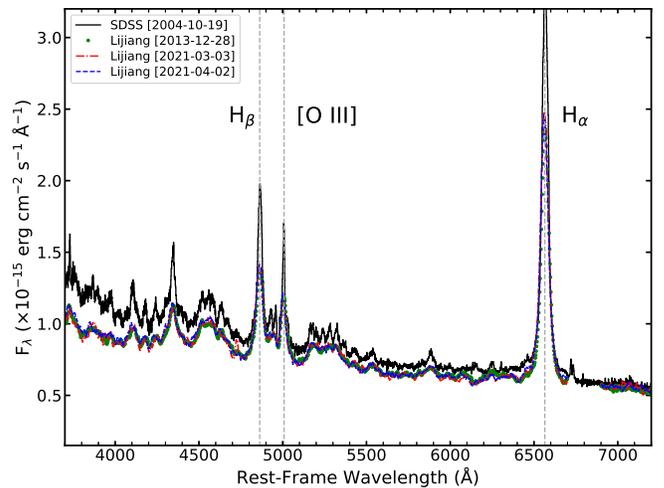}
\caption{The SDSS and Lijiang spectra of SDSS \hbox{J$0814+5325$}.
All spectra have been corrected for the Galactic extinction and smoothed with a boxcar of width 5 pixels.
The Lijiang spectra in the \hbox{rest-frame} \hbox{6700--6900 \AA} range are affected by strong telluric absorption and are not shown here.
The SDSS spectrum appears slightly brighter (by $\approx18\%$) than the Lijiang spectra below \hbox{rest-frame} $5100~{\rm \AA}$.
These spectra indicate that the accretion rate of SDSS \hbox{J$0814+5325$} did not change greatly over the years.
}
\label{fig:opt_spec}
\end{figure}

\subsection{Optical Spectra and Multiwavelength Photometric Data} \label{sec:2.4}

SDSS \hbox{J$0814+5325$} was reverberation mapped by the
Lijiang \hbox{2.4~m} telescope at the Yunnan Observatories of the Chinese Academy of Sciences between 2013 November and 2014 April \citep{Du2015}.
On 2021 March 3 and April 2, we obtained two additional Lijiang spectra that are nearly simultaneous to the \chandra\ and \xmm\ observations.
The data were reduced the same way as in \cite{Du2015}.
For comparison, the two latest Lijiang spectra, a typical Lijiang \hbox{reverberation-mapping} spectrum from 2013 December 28, and the SDSS spectrum observed on 2004 October 19 are shown in Figure~\ref{fig:opt_spec}.
The Lijiang spectra are consistent with each other overall.
The SDSS spectrum appears slightly brighter (by $\approx18\%$) than the Lijiang spectra below \hbox{rest-frame} $5100~{\rm \AA}$.
The mild difference between the spectra indicates that the accretion rate of SDSS \hbox{J$0814+5325$} did not change greatly over the years (see also Section~\ref{sec:3.4} below).

To investigate optical and \hbox{mid-infrared} variability of SDSS \hbox{J$0814+5325$}, we collected its \hbox{multi-epoch} multiwavelength measurements from the Near-Earth Object Wide-field Infrared Survey Explorer Reactivation Mission (\hbox{NEOWISE}, \citealt{Mainzer2011}), Catalina Real-Time Transient Survey (CRTS, \citealt{Drake2009}), Zwicky Transient Facility (ZTF, \citealt{Masci2019}), and Panoramic Survey Telescope and Rapid Response System (\hbox{Pan-STARRS}, \citealt{Flewelling2020}) catalogs.
The light curves and the spectral energy distribution (SED) are presented in Section~\ref{sec:3.4} below.

\begin{figure*}
\leftline{
\includegraphics[scale=0.33]{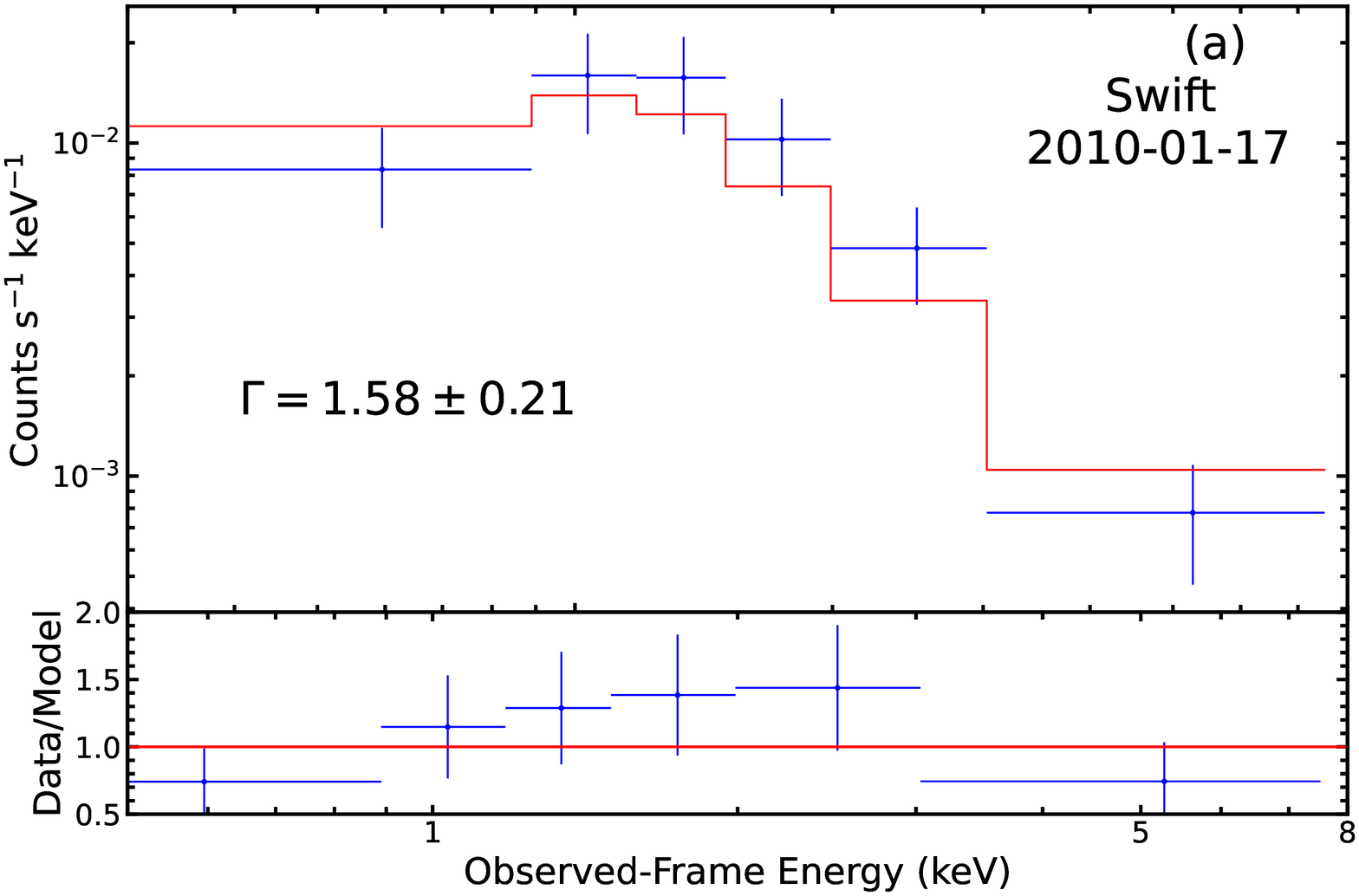}
\includegraphics[scale=0.33]{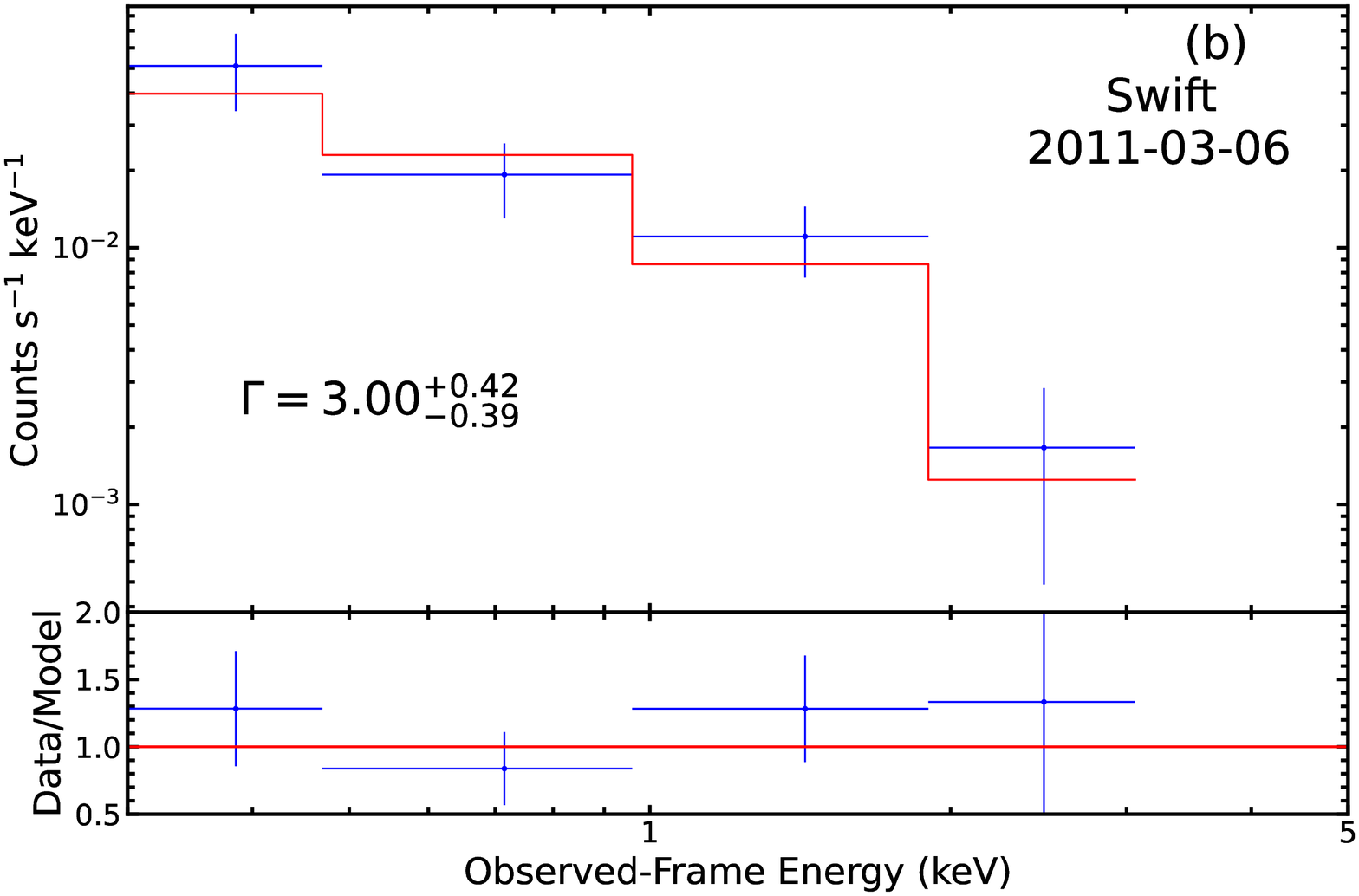}
}
\centerline{}
\leftline{
\includegraphics[scale=0.33]{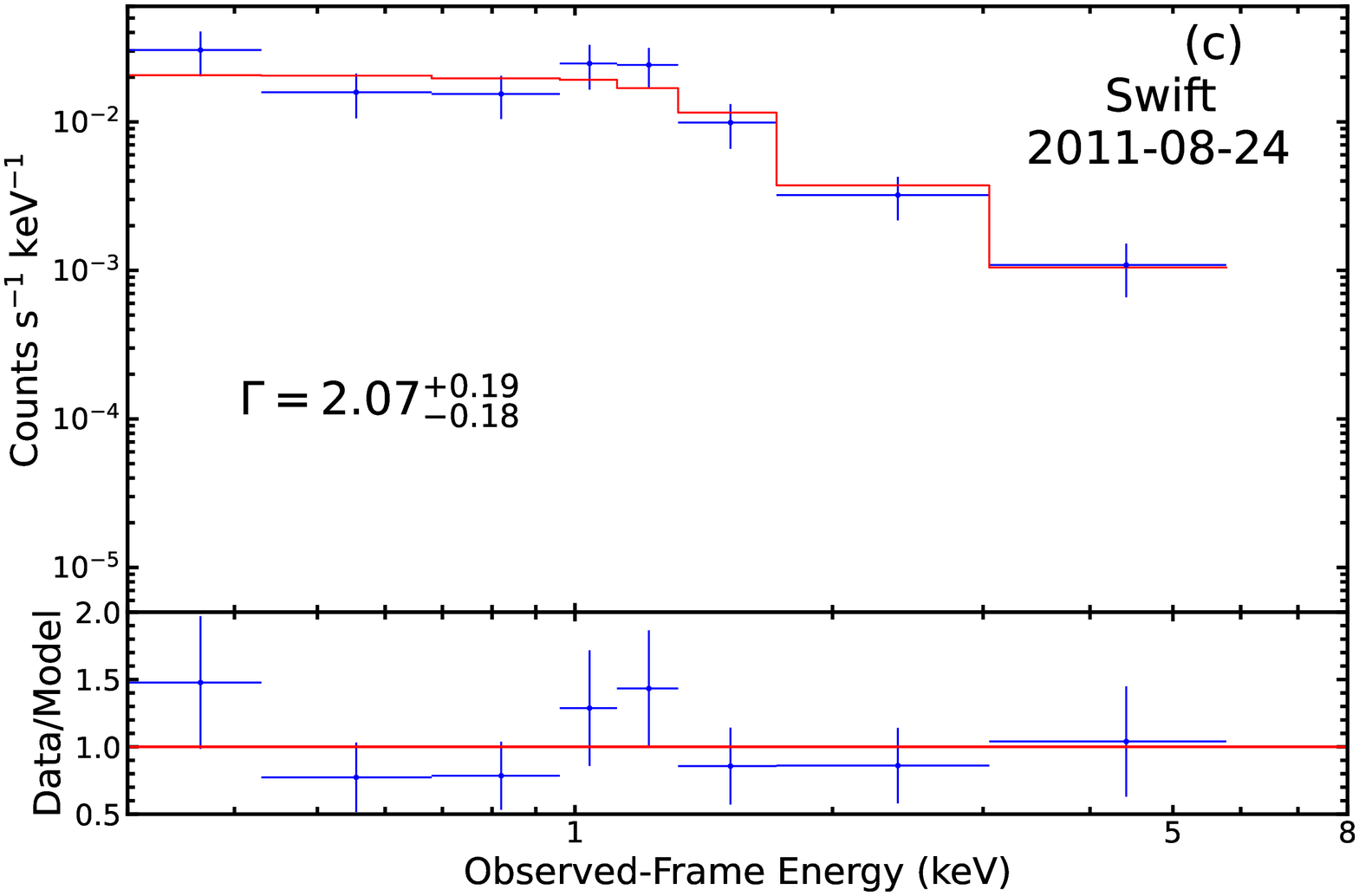}
\includegraphics[scale=0.33]{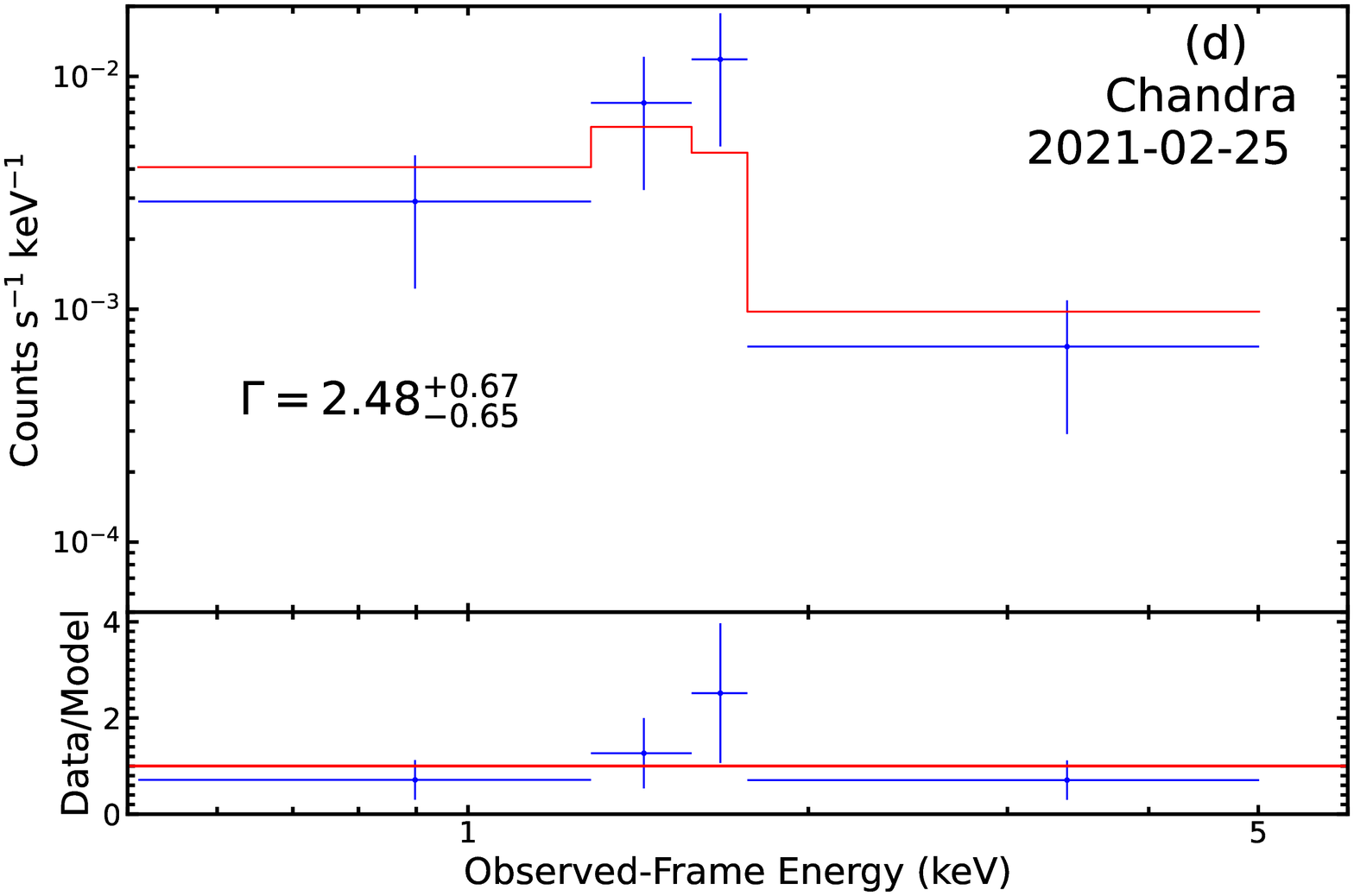}
}
\caption{The \xray\ spectra of the (a) 2010 January 17 \swift, (b) 2011 March 6 \swift, (c) 2011 August 24 \swift, and (d) 2021 February 25 \chandra\ observations, overlaid with the \hbox{best-fit} simple \hbox{power-law} model modified with Galactic absorption.
For display purposes, we grouped the spectra so that each bin has at least $3\sigma$ significance for the \swift\ spectra and $1.5\sigma$ significance for the \chandra\ spectrum.
The \hbox{best-fit} models are shown in red and the spectra are shown in blue.
The bottom panels show the ratios of the spectral data to the \hbox{best-fit} models.
}
\label{fig:cha_swi_spec}
\end{figure*}

\begin{figure*}
\centerline{
\includegraphics[scale=0.33]{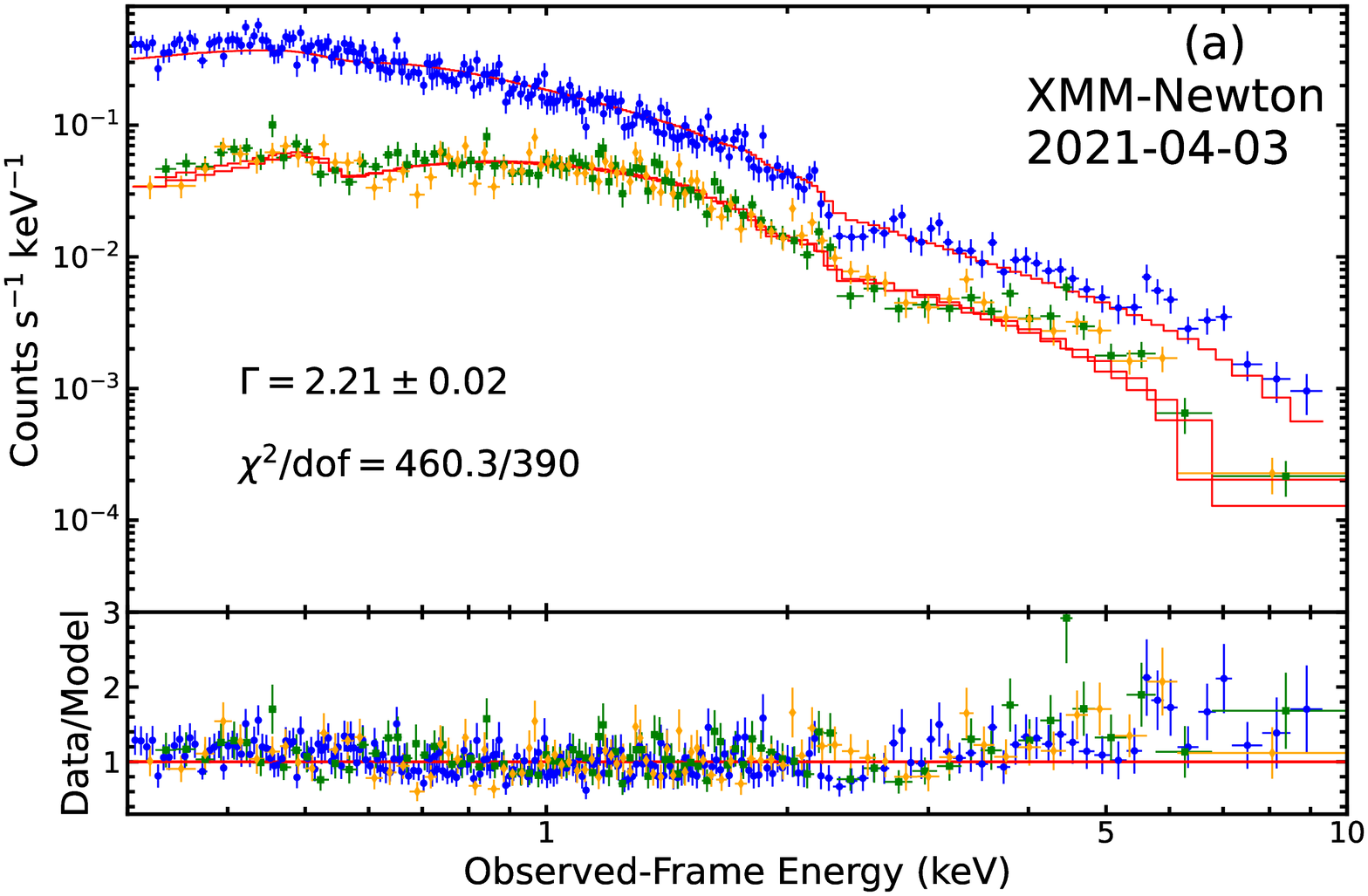}
\includegraphics[scale=0.33]{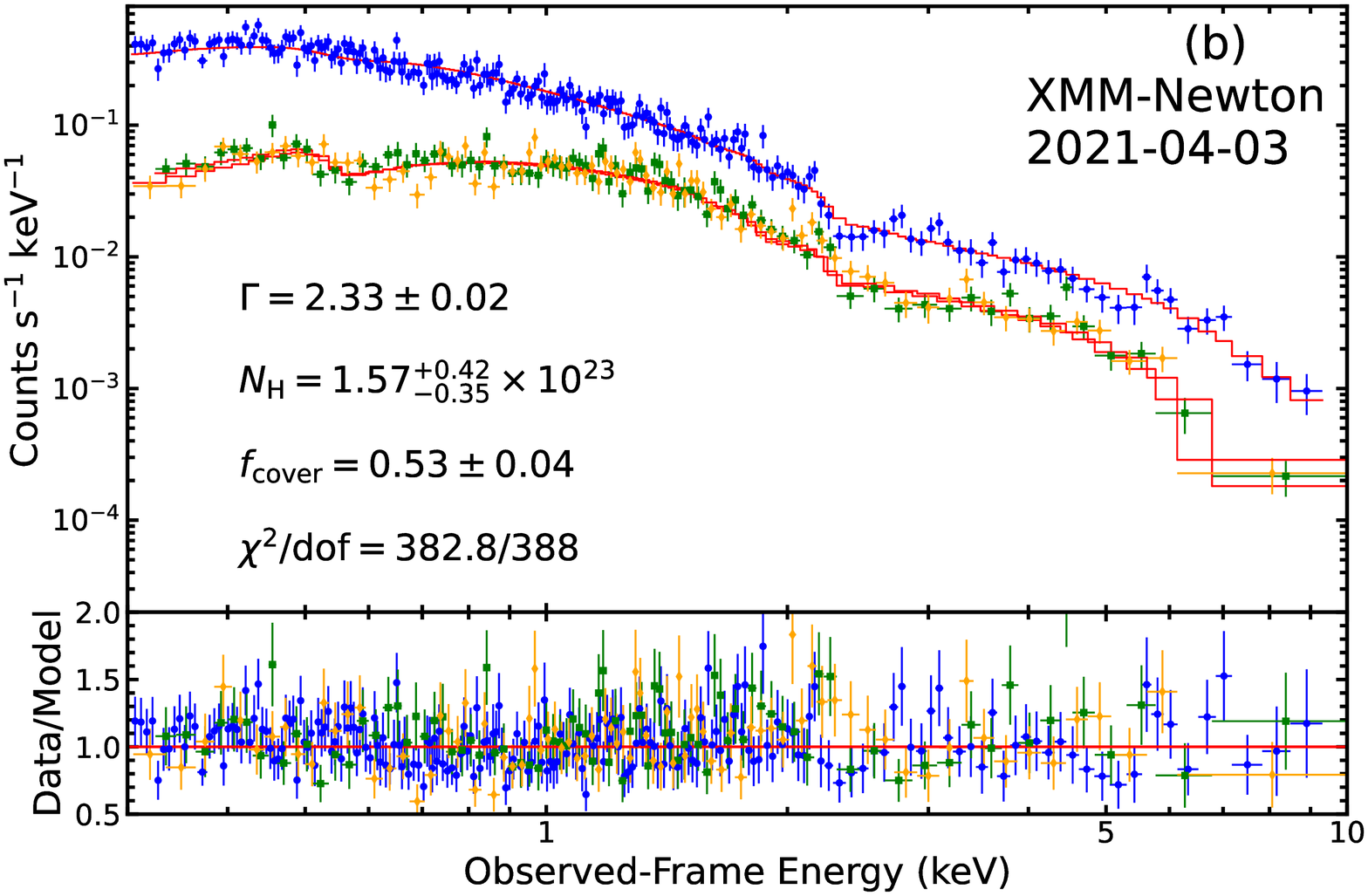}
}
\caption{The \xmm\ spectra overlaid with (a) the \hbox{best-fit} simple \hbox{power-law} model or (b) the \hbox{partial-covering} absorption model.
The spectra were grouped with at least 25 counts per bin.
The \hbox{best-fit} models are shown in red.
The \xmm\ pn, MOS1, and MOS2 spectra are shown in blue, green, and orange, respectively.
The bottom panels show the ratios of the spectral data to the \hbox{best-fit} models.
}
\label{fig:xray_xmm}
\end{figure*}

\section{X-ray and Multiwavelength Properties} 
\label{sec:3}

\subsection{X-ray Spectral Analysis} 
\label{sec:3.1}
We used XSPEC (v.12.11.0; \citealt{Arnaud1996}) to fit the \chandra, \xmm, and \swift\ spectra of SDSS \hbox{J$0814+5325$}.
The three \xmm\ spectra (from pn, MOS1, and MOS2) are jointly fitted.
The energy ranges used in the spectral fitting are \hbox{0.3--10 keV} for the \xmm\ spectra and \hbox{0.3--8 keV} for the \chandra\ and \swift\ spectra.
The $W$ statistic in XSPEC was used for the \chandra\ and \swift\ spectral fitting, and the $\chi^{2}$ statistic was used for the \xmm\ spectral fitting.
We first adopted a simple \hbox{power-law} model modified by Galactic absorption ({\sc zpowerlw*phabs}) for all the spectra.
The Galactic neutral hydrogen column density $N_{\rm H}$ was fixed at $3.59\times10^{20}~{\rm cm}^{-2}$ \citep{HI4PI2016}.
The \hbox{best-fit} results are displayed in Table~\ref{tbl:fitting} and Figures~\ref{fig:cha_swi_spec} and~\ref{fig:xray_xmm}.

The simple \hbox{power-law} model describes well the \chandra\ and \swift\ spectra overall (Figures~\ref{fig:cha_swi_spec}~and $W$/dof values in Table~\ref{tbl:fitting}), considering the limited photon statistics of these spectra.
The \hbox{best-fit} photon indices for the \chandra\ spectrum and the two 2011 \swift\ spectra are all $\gtrsim2$, consistent with the expectation for a \hbox{super-Eddington} accreting AGN \citep[e.g.,][]{Shemmer2008,Huang2020,Liu2021}.
For the 2010 \swift\ spectrum, the photon index ($1.58\pm{0.21}$) appears slightly smaller, and thus we added an intrinsic absorption component ({\sc zphabs}) to fit the spectrum.
The \hbox{best-fit} results are also shown in Table~\ref{tbl:fitting}.
The intrinsic absorption model significantly improves the fitting with $\Delta{W}/{\rm dof=3.9/1}$ and an \hbox{$F$-test} probability of $0.024$, and the resulting photon index ($2.26^{+0.46}_{-0.42}$) became consistent with those from the other spectra.
The intrinsic absorption appears mild with $N_{\rm H}\approx2.5\times10^{21}~{\rm cm}^{-2}$.

The \xmm\ spectra have good photon statistics, and significant residuals are apparent at $>5~{\rm keV}$ energies for the simple \hbox{power-law} fitting (Figure~\ref{fig:xray_xmm}a).
We then fitted only the \hbox{2--10 keV} spectra with the \hbox{power-law} model, and a smaller photon index ($1.78\pm{0.07}$) than the \hbox{broad-band} value ($2.20\pm{0.02}$) was obtained.
Such a hard \xray\ excess in an overall steep spectrum can be typically explained with \hbox{partial-covering} absorption.
We thus tested a \hbox{partial-covering} absorption model ({\sc phabs*zpcf*zpowerlw}) to fit the spectra, and the results are shown in Table~\ref{tbl:fitting} and Figure~\ref{fig:xray_xmm}b.
This model improves significantly the spectral fitting with $\Delta\chi^{2}/{\rm dof=77.5/2}$ and an \hbox{$F$-test} probability of \hbox{$2.87\times10^{-16}$}.
We also obtained a \hbox{null-hypothesis} probability (the probability of the observed data being drawn from the model) of $0.565$, which is significantly larger than that ($0.008$) obtained from the simple \hbox{power-law} model.
The resulting absorber column density is $\approx1.6\times10^{23}~{\rm cm}^{-2}$ with a covering factor of $\approx0.53$.

From the \hbox{best-fit} models, we computed the observed (not corrected for any intrinsic absorption) \hbox{2 keV} flux densities that are used later for the \aox\ computations.
The absorption models are used for the 2010 \swift\ spectrum and the \xmm\ spectra as these provide improved fits.
We also computed the observed \xray\ luminosities in the \hbox{rest-frame} \hbox{2--10} keV band.
These results are listed in Table~\ref{tbl:fitting}.

\begin{longrotatetable}
\begin{deluxetable*}{lcccccrcccccccc}
\tablecolumns{14}
\tabletypesize{\scriptsize}
\tablecaption{\xray\ Spectral Fitting Results}
\tablehead{
\colhead{Observatory}  & 
\colhead{Observation}  & 
\colhead{Spectral} & 
\colhead{$\Gamma$} & 
\colhead{$N_{\rm H}$} & 
\colhead{$f_{\rm cover}$} & 
\colhead{$W$/dof} & 
\colhead{$f_{\rm 2~keV}$} & 
\colhead{$L_{\rm X}$} & 
\colhead{$\alpha_{\rm OX}$} & 
\colhead{$\Delta\alpha_{\rm OX}$} & 
\colhead{$f_{\rm weak}$} & 
\colhead{$\alpha_{\rm OX, ~corr}$} & 
\colhead{$\Delta\alpha_{\rm OX, ~corr}$}\\ 
\colhead{ }   &
\colhead{Start Date}   &
\colhead{Counts}   &
\colhead{ } &
\colhead{$(10^{22}~{\rm cm}^{-2})$} &
\colhead{ } &
\colhead{($\chi^{2}$/dof)}   &
\colhead{ } &
\colhead{($10^{43}~{\rm erg}~~{\rm s}^{-1}$)}   &
\colhead{ }   &
\colhead{ }   &
\colhead{ }   &
\colhead{ }   &
\colhead{ }\\  
\colhead{(1)}   &
\colhead{(2)}   &
\colhead{(3)}   &
\colhead{(4)}   &
\colhead{(5)}   &
\colhead{(6)}   &
\colhead{(7)}   &
\colhead{(8)}   &
\colhead{(9)}   &
\colhead{(10)}  &
\colhead{(11)} &
\colhead{(12)} &
\colhead{(13)} &
\colhead{(14)}
}
\startdata
\multirow{2}{*}{Swift} & \multirow{2}{*}{2010 Jan 17} & \multirow{2}{*}{55} & $1.58^{+0.21}_{-0.21}$ &... &...& 39.2/50 & ...& ...&...&...&...&...&...\\
 &  &  & $2.26^{+0.46}_{-0.42}$ &$0.25^{+0.16}_{-0.13}$ &...& 35.3/49 & $10.61_{-2.85}^{+1.33}$ & $2.34_{-0.86}^{+0.61}$ &$-1.39_{-0.05}^{+0.02}$&$-0.03$&1.2&$-1.37_{-0.10}^{+0.08}$&$-0.01$\\
\multirow{2}{*}{Swift} & 2011 Mar 06 & 31 & $3.00^{+0.42}_{-0.39}$ &... &... & 29.3/26& $~4.31_{-1.36}^{+1.63}$ & $0.54_{-0.24}^{+0.37}$
&$-1.54_{-0.06}^{+0.05}$&$-0.18$&2.9&...&...\\
 & 2011 Aug 24  & 74 & $2.07^{+0.19}_{-0.18}$ &... &... & 62.3/58& $10.76_{-1.84}^{+2.03}$ & $2.51_{-0.65}^{+0.88}$
&$-1.39_{-0.03}^{+0.03}$&$-0.03$&1.2&...&...\\
{Chandra} & 2021 Feb 25 & 12 & $2.48^{+0.67}_{-0.65}$ &... &... & 12.7/9& $~1.12_{-0.61}^{+0.90}$ & $0.19_{-0.14}^{+0.33}$
&$-1.77_{-0.13}^{+0.10}$&$-0.41$&11.7&...&...\\
\multirow{2}{*}{XMM-Newton}& \multirow{2}{*}{2021 Apr 03} & \multirow{2}{*}{11941} & $2.21^{+0.02}_{-0.02}$ & ... &...& (460.3/390)&...&...&...&...&...&...&...\\
 & &  & $2.32^{+0.02}_{-0.02}$& $15.74_{-3.50}^{+4.25}$ & $0.53_{-0.04}^{+0.04}$& (382.8/388)&$~5.97_{-0.13}^{+0.12}$ & $1.64_{-0.01}^{+0.01}$ &$-1.49_{-0.01}^{+0.01}$&$-0.13$&2.2&$-1.37_{-0.01}^{+0.01}$&$-0.01$
\enddata
\tablecomments{
Column (1): \xray\ observatory.
Column (2): observation start date.
Column (3): net spectral counts.
Column (4): \xray\ photon index.
Column (5): column density of the \xray\ absorber if an absorption model is adopted.
Column (6): covering factor of the \xray\ absorber if a \hbox{partial-covering} absorption model is adopted.
Column (7): $W$ (or $\chi^{2}$) statistic value over the degrees of freedom.
Column (8): flux density at \hbox{rest-frame} \hbox{2 keV} in units of $10^{-31}~{\rm erg}~{\rm cm}^{-2}~{\rm s}^{-1}~{\rm Hz}^{-1}$.
Column (9): luminosity in the \hbox{rest-frame} \hbox{2--10 keV} band.
Columns (10)--(11): observed \aox\ value, and its deviation from the \aox\ value expected from the \cite{Steffen2006} \hbox{\aox--$L_{2500~\textup{\AA}}$} relation.
Column (12): \xray\ weakness parameter (see Section~\ref{sec:3.3}).
Columns (13)--(14): \hbox{absorption-corrected} \aox\ value, and its deviation from the \aox\ value expected from the \cite{Steffen2006} \hbox{\aox--$L_{2500~\textup{\AA}}$} relation.
}
\label{tbl:fitting}
\end{deluxetable*}
\end{longrotatetable}

\begin{figure}
\centering
\includegraphics[scale=0.33]{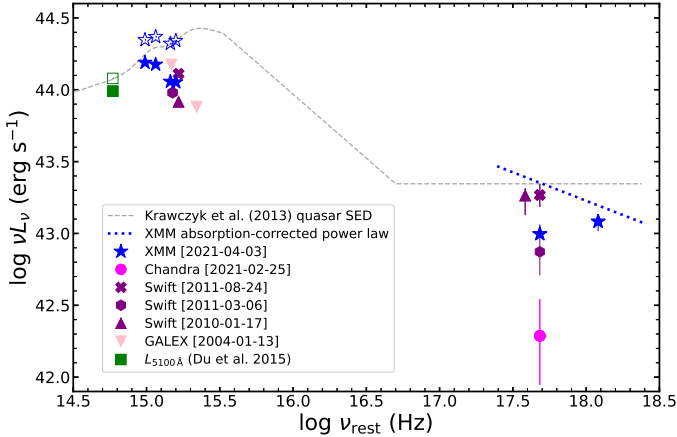}
\caption{Optical-to-\xray\ SED of SDSS \hbox{J$0814+5325$}.
The \xmm\ OM data points are shown as the solid blue stars.
The rightmost OM data point (UVW2) and the \hbox{5100~\AA} AGN luminosity (green square) estimated by \cite{Du2015} were used to derive an intrinsic $E(B-V)$ of 0.073 by comparing these measurements to the \cite{Krawczyk2013} mean quasar SED.
The \hbox{extinction-corrected} 5100 \AA\ luminosity and the OM measurements are shown as the open square and stars, and the expected intrinsic SED is shown as the gray dashed curve.
We also show the \swift\ UVOT and \galex\ UV photometric data, which indicate intrinsic extinction as well.
The observed \hbox{2 keV} luminosities from the \hbox{best-fit} models are displayed for comparison with the expected \xray\ luminosity (gray dashed line in the \xray\ energies) from the \cite{Steffen2006} \aox--$L_{2500~\textup{\AA}}$ relation.
The 2010 \swift\ data point is displaced slightly along the \hbox{X-axis} for display purposes.
For the \xmm\ observation, we also show the observed \hbox{5 keV} luminosity and the intrinsic $\Gamma=2.37$ \hbox{power-law} continuum (blue dotted line) from the \hbox{best-fit} model.
}
\label{fig:sed_dered}
\end{figure}

\begin{figure}
\centering
\includegraphics[scale=0.33]{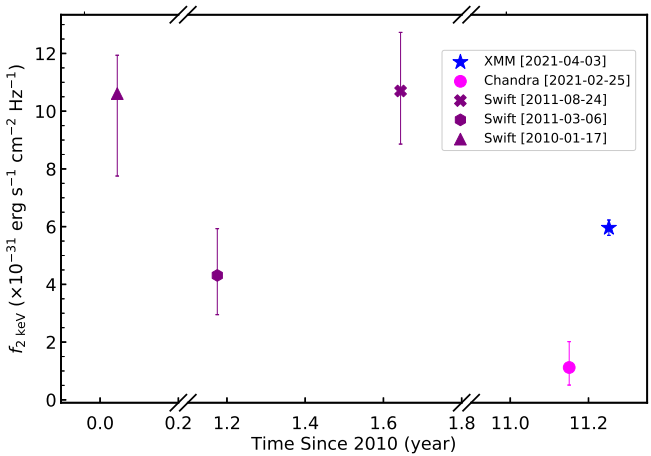}
\caption{
The $f_{\rm 2~keV}$ light curve of SDSS \hbox{J0814+5325}.
}
\label{fig:ftkev}
\end{figure}

\subsection{UV Extinction Correction for the \xmm\ Observation} \label{sec:3.2}

We show the \xmm\ OM measurements (UVW2, UVM2, UVW1, and U bands) in Figure~\ref{fig:sed_dered}.
The mean SED of low luminosity quasars in \cite{Krawczyk2013} is shown for comparison.
The UV SED shape differs significantly from that of typical quasars, and a natural explanation is intrinsic extinction.
It is difficult to estimate the extinction directly as there is likely \hbox{host-galaxy} contamination, at least in the UVW1 and U bands.
Since there is a simultaneous Lijiang spectrum, which is also consistent with the Lijiang \hbox{reverberation-mapping} spectra, we adopted the \hbox{5100~\AA} AGN luminosity estimate (${\rm log}[L_{5100~{\textup{\AA}}}/{\rm erg}~{\rm s}^{-1}]=43.99$) in \cite{Du2015}, which excludes the \hbox{host-galaxy} contamination of ${\rm log}(L_{5100~{\textup{\AA}},~{\rm host}}/{\rm erg}~{\rm s}^{-1})=43.73$ using Equation 1 of \cite{Shen2011}.
This \hbox{5100~\AA} AGN luminosity is also shown in Figure~\ref{fig:sed_dered}.
Assuming that the UVW2 (\hbox{2120~\AA}) measurement has negligible \hbox{host-galaxy} contamination, we estimated the extinction by comparing the UVW2 and \hbox{5100~\AA} measurements to the \cite{Krawczyk2013} mean SED.
We adopted the Small Magellanic Cloud (SMC) extinction model (\citealt{Gordon2003}; $R_V=2.74$), which is commonly used to model the intrinsic extinction of AGNs \citep[e.g.,][]{Hines&Wills1995,Hopkins2004,Glikman2012}.
An $E(B-V)$ value of 0.073 was obtained, and the \hbox{extinction-corrected} \hbox{5100~\AA} and UV luminosities are shown in Figure~\ref{fig:sed_dered}.
By definition, the corrected \hbox{5100~\AA} and the UVW2 luminosities align with the \hbox{absorption-corrected} intrinsic quasar SED.
The corrected UVW1 and U luminosities are slightly higher than the SED due to \hbox{host-galaxy} contamination; the corrected UVM2 (\hbox{2310 \AA}) luminosity shows negligible excess, suggesting little \hbox{host-galaxy} contamination at this wavelength.

In Figure~\ref{fig:sed_dered}, we show also the \swift\ UVOT and Galaxy Evolution Explorer (\galex; \citealt{Martin2005}) UV photometric data points.
The SED shape from the \galex\ FUV (1528~\AA) and NUV (2271~\AA) measurements indicate extinction of a similar level ($E(B-V)\approx 0.15$).
The \swift\ UVOT used only one filter per observation and thus we cannot constrain an extinction value.
However, the significant deviations of the measurements from the expected intrinsic SED suggest intrinsic extinction as well.

Given the UV extinction estimation for the \xmm\ observation, we derived the corresponding neutral hydrogen column density adopting the standard Galactic \hbox{gas-to-dust} ratio of \cite{Rachford2009}:
\begin{equation}
N_{\rm H}=5.94\times10^{21}~E(B-V)~{\rm cm}^{-2}.
\end{equation}
The resulting column density of the UV absorber was about $4.3\times10^{20}~{\rm cm}^{-2}$.
We verified that adding such an \xray\ absorber to our \xmm\ spectral modeling in Section~\ref{sec:3.1} does not affect significantly the \hbox{best-fit} parameters.
The \swift\ spectra can also accommodate similar \xray\ absorbers considering the limited quality of the spectra.


\subsection{Strong and Rapid X-ray Variability and the \aox\ Parameter} \label{sec:3.3}

We plot the $f_{\rm 2~keV}$ light curve of SDSS \hbox{J$0814+5325$} in Figure~\ref{fig:ftkev}, showing strong and rapid \xray\ variability.
The largest variability amplitude was observed between the 2011 August \swift\ observation and the 2021 \chandra\ observation, being $9.6^{+11.6}_{-4.6}$.
The $1\sigma$ uncertainties of the variability amplitude were derived from that of $f_{\rm 2~keV}$ following the method described in Section 1.7.3 of \cite{Lyons1991}.
The \hbox{2--10 keV} luminosity varied by a factor of $13.2^{+37.3}_{-9.1}$ between the two observations (Table~\ref{tbl:fitting}).
Between the \chandra\ and \xmm\ observations, the \hbox{rest-frame} \hbox{2 keV} flux density varied by a factor of $5.3^{+6.4}_{-2.4}$ in \hbox{rest-frame} $32$ days. 

\begin{figure}
\leftline{
\includegraphics[scale=0.33]{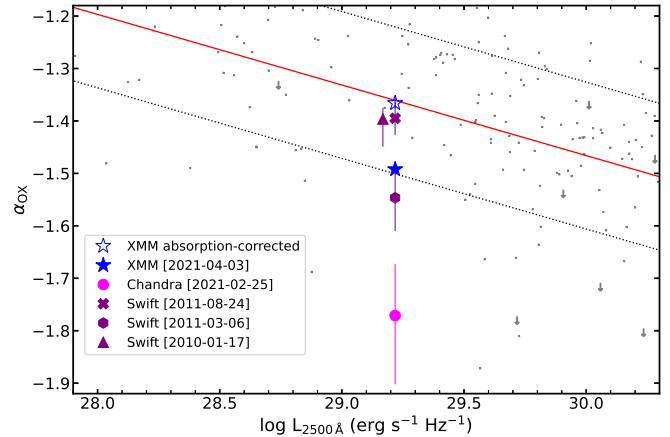}
}
\caption{X-ray-to-optical power-law slope ($\alpha_{\rm OX}$) vs. 2500 \AA\ luminosity for the \xray\ observations of SDSS \hbox{J$0814+5325$}. 
The $f_{\rm 2500~\textup{\AA}}$ value computed from the \hbox{extinction-corrected} SED in Figure~\ref{fig:sed_dered} was adopted for all the measurements; the 2010 \swift\ data point is displaced slightly along the \hbox{X-axis} for display purposes.
We also show the \hbox{absorption-corrected} $\alpha_{\rm OX}$ value for the \xmm\ observation (blue open star).
The gray dots and downward arrows (upper limits) represent the $\alpha_{\rm OX}$ values of the \cite{Steffen2006} sample.
The red solid line shows the \cite{Steffen2006} $\alpha_{\rm OX}$--$L_{\rm 2500~{\textup{\AA}}}$ relation, and the black dotted lines show the $1\sigma$ scatter ($\approx0.14$) of this relation.
}
\label{fig:aox}
\end{figure}

We then computed the \aox\ value for each \xray\ observation.
For the \xmm\ observation, we used the $f_{2500~\textup{\AA}}$ value ($4.61\times10^{-27}~{\rm erg~cm^{-2}~s^{-1}~Hz^{-1}}$) calculated from the \hbox{extinction-corrected} SED in Figure~\ref{fig:sed_dered}.
The \chandra\ observation is only 32 days apart and thus we adopted the same 
$f_{2500~\textup{\AA}}$ value.
The \swift\ UVOT measurements are likely affected by uncertain intrinsic extinction (Section~\ref{sec:3.2}), and thus simultaneous $f_{2500~\textup{\AA}}$ measurements are not available.
Considering that the \hbox{long-term} optical variability appears much weaker than the \xray\ variability (see Section~\ref{sec:3.4} below), we still adopted the same $f_{2500~\textup{\AA}}$ value for the \aox\ computations.
The resulting \aox\ values are listed in Table~\ref{tbl:fitting}, and the \aox\ versus $L_{2500~\textup{\AA}}$ distribution is shown in Figure~\ref{fig:aox}.
For comparison, the \cite{Steffen2006} $\alpha_{\rm OX}$--$L_{2500~\textup{\AA}}$ relation and its $1\sigma$ scatter are also displayed in Figure~\ref{fig:aox}.
For the 2010 January and 2011 August Swift observations, SDSS \hbox{J$0814+5325$} showed \hbox{nominal-strength} \xray\ emission, while for the other observations, its \xray\ emission was weaker than that expected from the $\alpha_{\rm OX}$--$L_{2500~\textup{\AA}}$ relation. 
We quantified the level of \xray\ weakness by computing the $\Delta\alpha_{\rm OX}$ parameter, defined as the difference between the observed and expected $\alpha_{\rm OX}$ values ($\Delta\alpha_{\rm OX}$=$\alpha_{\rm OX}-\alpha_{\rm OX,~exp}$).
The corresponding \xray\ weakness factor at \hbox{2 keV} is $f_{\rm weak}=403^{-\Delta\alpha_{\rm OX}}$.
The $\Delta\alpha_{\rm OX}$ and $f_{\rm weak}$ values are listed in Table~\ref{tbl:fitting}.
The \chandra\ observation displays the weakest \xray\ emission with $\Delta\alpha_{\rm OX}=-0.41_{-0.13}^{+0.10}$ and $f_{\rm weak}=11.7_{-6.3}^{+9.6}$, and the 2011 March \swift\ observation also displays weak \xray\ emission with $\Delta\alpha_{\rm OX}=-0.18_{-0.06}^{+0.05}$ and $f_{\rm weak}=2.9_{-1.3}^{+0.7}$.
These two spectra are fitted with simple \hbox{power-laws} with large photon indices (Table~\ref{tbl:fitting}).
The models describe well the spectra with no significant residuals (Figure~\ref{fig:cha_swi_spec}b and \ref{fig:cha_swi_spec}d).
We verified that adding an intrinsic absorption component does not improve the fits, and the \hbox{best-fit} intrinsic $N_{\rm H}$ values are low ($\approx{10}^{18}~{\rm cm}^{-2}$).
Considering also the large \hbox{best-fit} photon indices, these two spectra do not directly indicate any intrinsic absorption (but see Sections~\ref{sec:4.1} and \ref{sec:4.2} below).

For the 2010 \swift\ observation and the \xmm\ observation, absorption models were adopted to describe the spectra (Section~\ref{sec:3.1}). 
To probe the intrinsic \xray\ emission for these two observations, we computed the \hbox{absorption-corrected} $\alpha_{\rm OX}$ and $\Delta\alpha_{\rm OX}$ values from the \hbox{absorption-corrected} $f_{\rm 2~keV}$ values. 
The results are listed in Table~\ref{tbl:fitting}. 
The absorption correction is small for the \swift\ observation, with $\Delta\alpha_{\rm OX}=-0.03$ and $\Delta\alpha_{\rm OX,~corr}=-0.01$, while it is more significant for the \xmm\ observation, with $\Delta\alpha_{\rm OX}=-0.13$ and $\Delta\alpha_{\rm OX,~corr}=-0.01$. 
This result suggests that SDSS \hbox{J$0814+5325$} still radiates a nominal level of \xray\ emission at the time of the \xmm\ observation, and the relatively weak \xmm\ emission observed is simply due to the \hbox{partial-covering} absorption described with the \hbox{best-fit} model. 
We also plotted the $\alpha_{\rm OX,~corr}$ value for the \xmm\ observation in Figure~\ref{fig:aox}.

\subsection{Optical--infrared Variability and Spectral Energy Distribution
} \label{sec:3.4}

\begin{figure}
\centering
\includegraphics[scale=0.38]{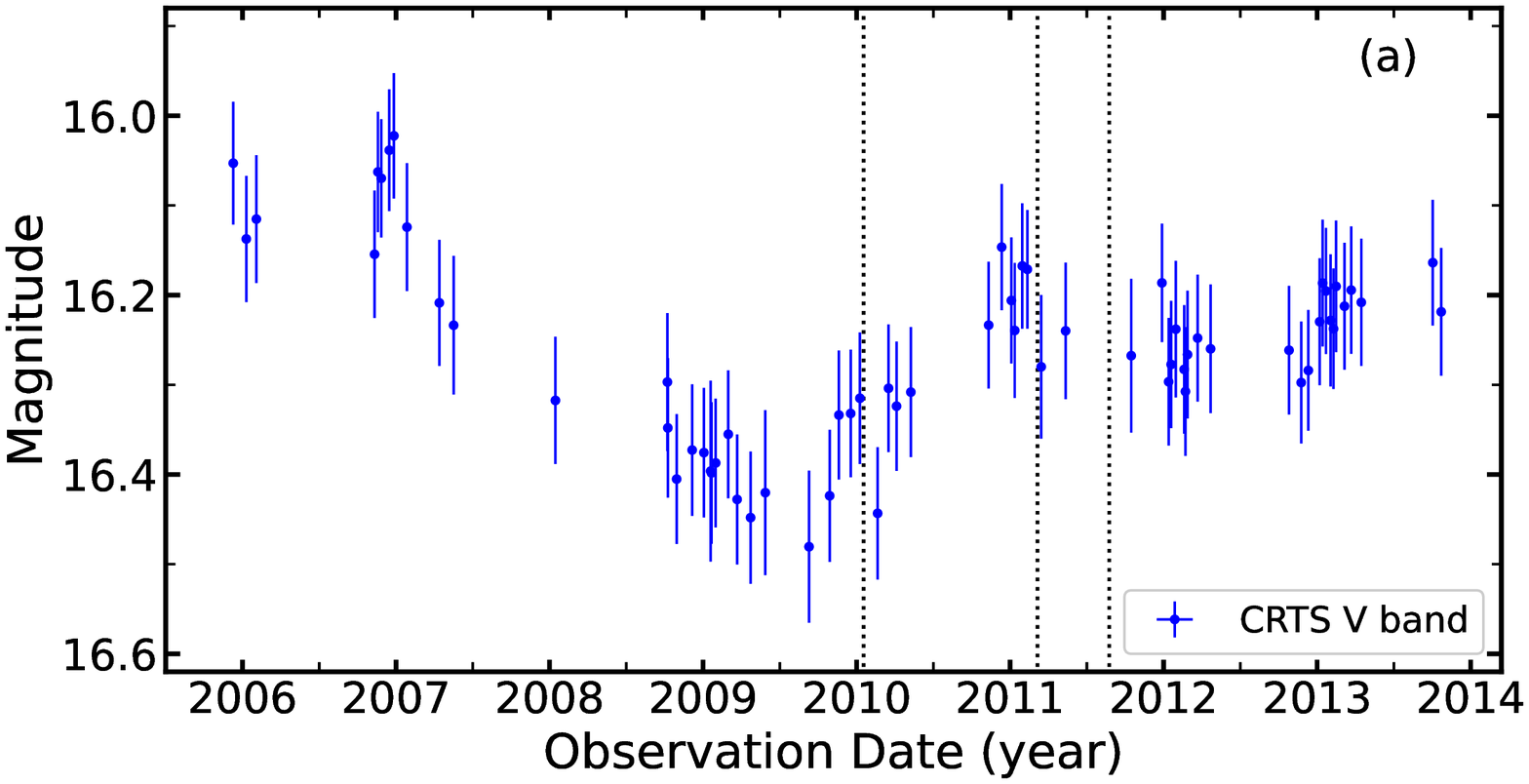}
\includegraphics[scale=0.38]{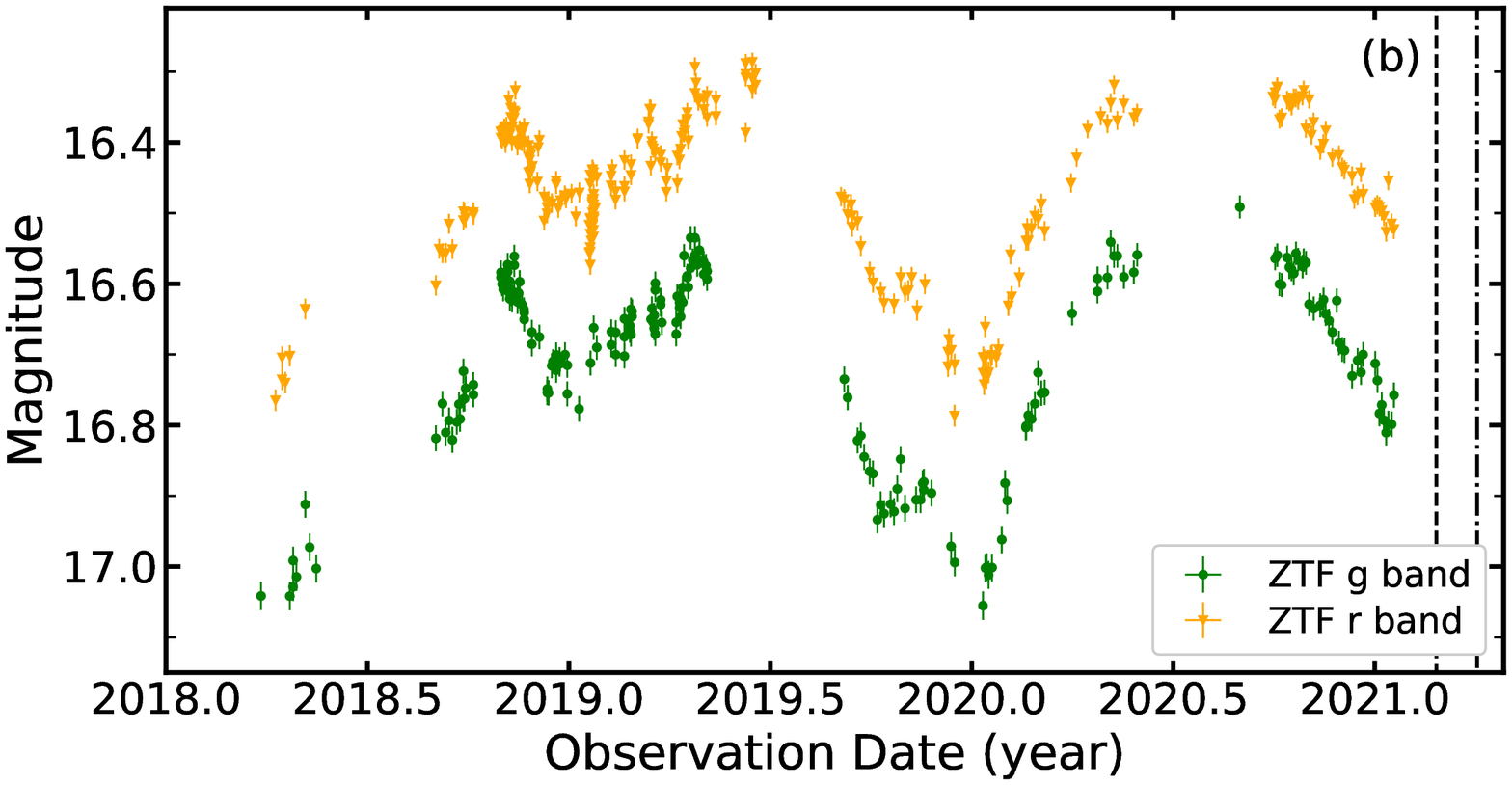}
\includegraphics[scale=0.38]{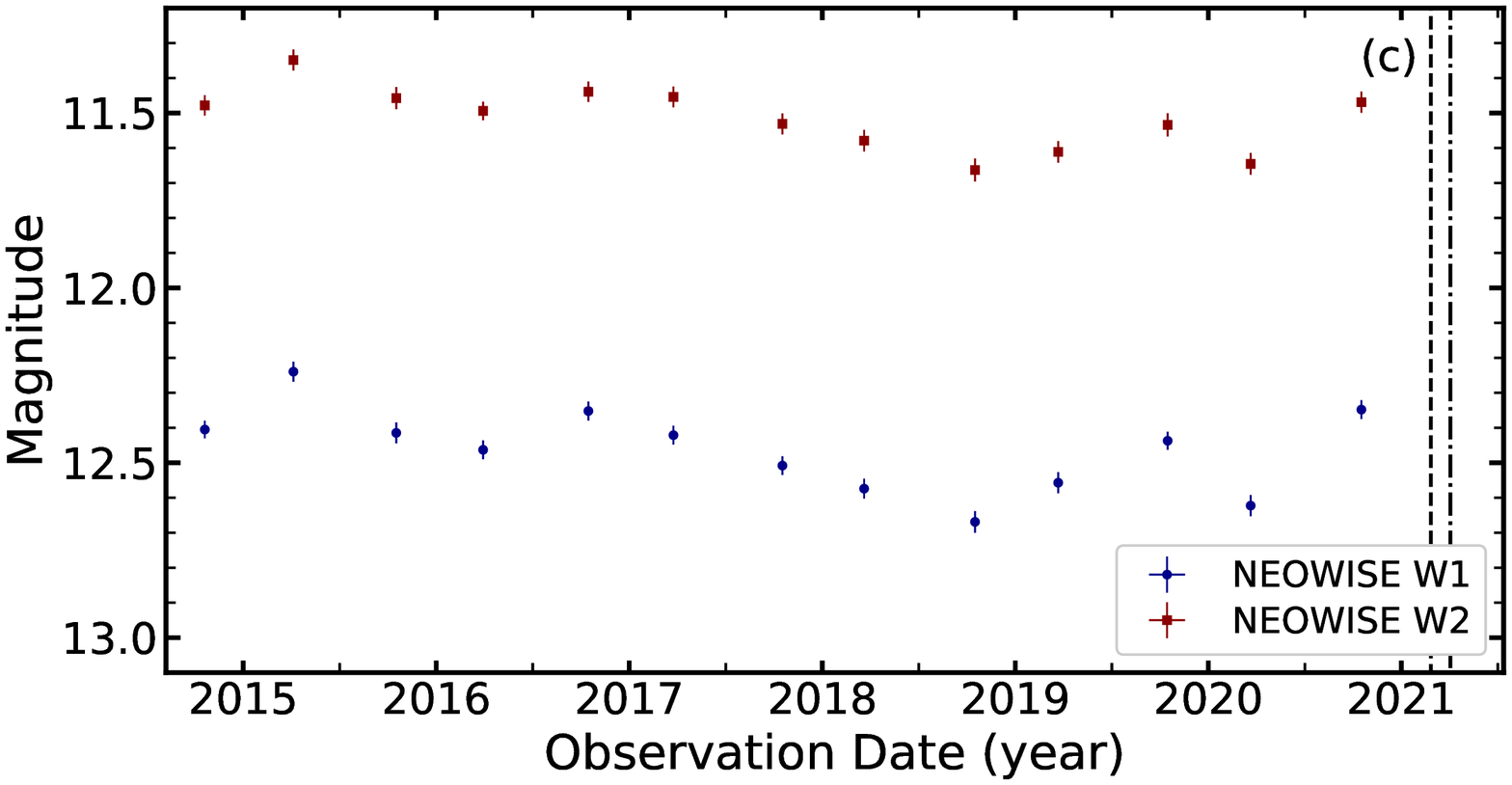}
\caption{
Light curves in the (a) CRTS $V$ band, (b) ZTF $g$ and $r$ bands, and (c) NEOWSIE W1 and W2 bands; we grouped \hbox{intra-day} measurements.
The vertical dotted, dashed, and \hbox{dashed-dotted} lines represent the dates of the \swift, \chandra, and \xmm\ observations, respectively.
These light curves indicate that SDSS \hbox{J0814+5325} does not show any substantial \hbox{long-term} variability in the IR and optical bands.
}
\label{fig:ztf_wise}
\end{figure}

We constructed optical--infrared (IR) light curves using the archival CRTS ($V$ band), ZTF ($g$ and $r$ bands), and NEOWISE (W1 and W2 bands) data, as shown in Figure~\ref{fig:ztf_wise}.
We do not present the \hbox{Pan-STARRS} optical light curves here as they contain limited numbers of photometric measurements, and the overall variability is consistent with that shown in the CRTS and ZTF light curves.
The maximum \hbox{long-term} optical variability amplitude reaches $\approx0.5~{\rm mag}$ ($\approx60\%$), but we caution that some of the variability might be caused by variable seeing conditions of the \hbox{ground-based} telescopes as the optical emission of SDSS \hbox{J$0814+5325$} has a strong extended \hbox{host-galaxy} component.
The mild optical variability is also consistent with the mild spectral variability shown in Figure~\ref{fig:opt_spec}.
In the NEOWISE light curves, the maximum variability amplitude reaches $\approx50\%$. 
A \hbox{long-term} max flux variability amplitude in the optical/IR bands of $\approx50\%$ is not unusual among AGNs, and the corresponding X-ray variability amplitude is only $\approx20\%$ given the $\alpha_{\rm OX}$--$L_{\rm 2500~{\textup{\AA}}}$ relation, insufficient to explain the observed strong \xray\ variability described in Section~\ref{sec:3.3}.

\begin{figure}
\centering
\includegraphics[scale=0.33]{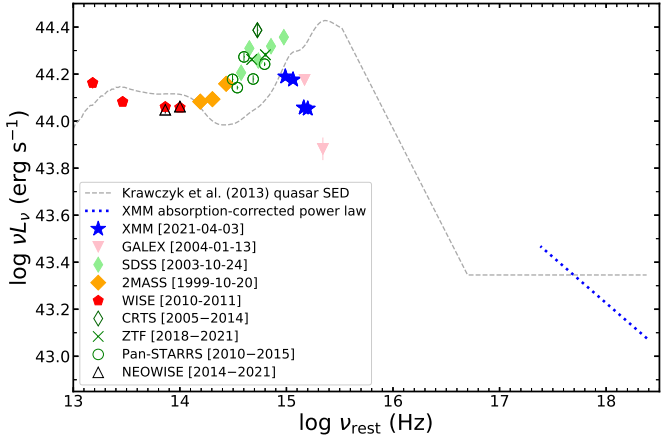}
\caption{
\hbox{IR-to-\xray} SED for SDSS \hbox{J$0814+5325$}.
The multiwavelength survey names and observational dates/periods are listed.
For the CRTS, ZTF, Pan-STARRS, and NEOWISE surveys, the average measurements of \hbox{multi-epoch} observations are shown.
The dotted line is the \hbox{absorption-corrected} intrinsic \xray\ continuum derived from the \xmm\ spectral analysis.
The dashed line is the same as that in Figure~\ref{fig:sed_dered}, showing the expected intrinsic AGN SED.
The optical part of the SED has a significant \hbox{host-galaxy} contribution.
}
\label{fig:sed}
\end{figure}

We composed the \hbox{IR-to-X-ray} SED of SDSS \hbox{J$0814+5325$}, shown in Figure~\ref{fig:sed}.
The \hbox{IR-to-UV} data points were corrected for the Galactic extinction.
We collected the IR, optical, and UV measurements of SDSS \hbox{J$0814+5325$} from the {Wide-field~Infrared~Survey~Explorer} (WISE; \citealt{Wright2010}), Two Micron All Sky Survey (2MASS; \citealt{Skrutskie2006}), SDSS, and GALEX catalogs.
The GALEX data presented in this paper were obtained from the Mikulski Archive for Space Telescopes (MAST) at the Space Telescope Science Institute. The specific observations analyzed can be accessed via \dataset[DOI: 10.17909/T9H59D]{https://doi.org/10.17909/T9H59D}.
We also show the average measurements from CRTS, ZTF, \hbox{Pan-STARRS}, and NEOWISE in Figure~\ref{fig:sed}.
These SED data are not simultaneous and they may be affected by variability effects with a maximum amplitude of $\approx50\%$.
The observed \xray\ data are displayed in Figure~\ref{fig:sed_dered} above, and in Figure~\ref{fig:sed} we show only the \hbox{absorption-corrected} intrinsic \xray\ continuum derived from the \xmm\ spectral analysis.
The expected intrinsic AGN SED inferred in Section~\ref{sec:3.2} is displayed for comparison.
The optical part of the SED appears to have a significant \hbox{host-galaxy} contribution, while the IR part is broadly consistent with the AGN SED.
Significant dust extinction is evident in the UV (as also discussed in Section~\ref{sec:3.2}).

\begin{deluxetable}{lcccc}
\tablewidth{0pt}
\tablecaption{Best-fit Properties of the \xray\ Absorber}
\tablehead{
\colhead{Observatory} &
\colhead{Observation} &
\colhead{$N_{\rm H}$*} &
\colhead{$f_{\rm cover}$} \\
\colhead{} &
\colhead{Start Date} &
\colhead{$(10^{22}~{\rm cm}^{-2})$} &
\colhead{ } \\
\colhead{(1)} &
\colhead{(2)} &
\colhead{(3)} &
\colhead{(4)} &   
}
\startdata
\multirow{3}{*}{Swift} & 2010 Jan 17 & $0.30_{-0.10}^{+0.24}$ & $0.96_{-0.16}^{+0.04}$ \\
 & 2011 Mar 06 & $18.13_{-14.11}$ & $0.45_{-0.12}^{+0.10}$  \\
 & 2011 Aug 24 & $0.80_{-0.57}^{+10.01}$ & $0.38_{-0.14}^{+0.17}$ \\
{Chandra} & 2021 Feb 25 & $70.95_{-34.44}$  & $0.91_{-0.03}^{+0.02}$ \\
{XMM-Newton}& 2021 Apr 03 & $15.74_{-3.50}^{+4.25}$ & $0.53_{-0.04}^{+0.04}$
\enddata
\tablecomments{
Column (1): \xray\ observatory.
Column (2): observation start date.
Column (3): column density of the \xray\ absorber.
Column (4): covering factor of the \xray\ absorber.
* No upper bound is available if the spectrum is not sensitive to very large $N_{\rm H}$ values.
}
\label{tbl:pcf}
\end{deluxetable}

\section{Discussion}
\label{sec:4}
\subsection{Obscuration Induced Rapid X-ray Variability}
\label{sec:4.1}

Given the lack of simultaneous optical spectral variability (Section~\ref{sec:2.4}), mild long-term optical/IR variability (Section~\ref{sec:3.4}), and the \xray\ spectral fitting results (Sections~\ref{sec:3.1} and~\ref{sec:3.3}), a natural explanation for the strong and rapid \xray\ variability in SDSS \hbox{J$0814+5325$} is a change of \xray\ obscuration from a \hbox{dust-free} absorber. 
Considering that it is a \hbox{super-Eddington} accreting quasar, the absorber is likely a \hbox{small-scale} clumpy \hbox{accretion-disk} wind\footnote{The absorber might also involve a thick inner accretion disk plus its associated outflow from \hbox{super-Eddington} accretion \citep[e.g.,][]{Luo2015,Ni2020,Ni2022}.}  that has been suggested to be responsible for similar \xray\ variability in other super-Eddington accreting quasars (Section~\ref{sec:intro}).
In this scenario, the intrinsic \xray\ continuum does not change in the short term (e.g., between the Chandra and XMM-Newton observations), and in the long term (year timescales) there is at most mild variability ($\approx20\%$ in accordance with the optical/IR variability) of the intrinsic \xray\ continuum.
The observed \xray\ variability is mainly due to changes of the column density and/or covering factor of the clumpy disk wind along the line of sight. This \xray\ absorber is \hbox{dust-free} and does not affect the observed optical/UV radiation.
The UV absorber inferred from the UV extinction in Section~\ref{sec:3.2} is located at a much larger scale; it is likely much more stable and does not cause significant optical variability.

Under the assumption that the intrinsic \xray\ continuum does not vary, we investigated the column densities and covering factors of the absorber among different \xray\ observations.
We consider the intrinsic continuum to be the one inferred from the \xmm\ \hbox{spectral-fitting} results, and for the \chandra\ and \swift\ observations, we fix this continuum (i.e., $\Gamma$ and normalization) and fit for the absorber parameters with the \hbox{partial-covering} absorption model.
This model explains well the \xray\ spectra in general, and the \hbox{best-fit} results are listed in Table~\ref{tbl:pcf}.
The 2010 \swift\ spectrum is described with mild obscuration with nearly full covering, consistent with the \hbox{best-fit} model in Table~\ref{tbl:fitting}.
The 2011 March \swift\ spectrum is affected by heavy (with no upper bound on $N_{\rm H}$) \hbox{partial-covering} absorption and the spectrum is dominated by the leaked component that is $\approx50\%$ of the intrinsic continuum.
The 2011 August \swift\ spectrum is described with mild \hbox{partial-covering} absorption, so that the spectrum is similar to the intrinsic continuum, showing a slightly smaller apparent $\Gamma$ ($2.07_{-0.18}^{+0.19}$ in Table~\ref{tbl:fitting}) and little X-ray weakness ($\Delta\alpha_{\rm OX}=-0.03$ in Table~\ref{tbl:fitting}).
The unusually weak \xray\ spectrum from the \chandra\ observation is described by heavy (with no upper bound on $N_{\rm H}$) obscuration with a $\approx91\%$ covering factor; the observed spectrum is dominated by the $\approx9\%$ leaked intrinsic continuum, and thus it is very weak ($\Delta\alpha_{\rm OX}=-0.41$ in Table~\ref{tbl:fitting}) yet it appears steep ($\Gamma=2.48_{-0.65}^{+0.67}$ in Table~\ref{tbl:fitting}) without absorption signatures.
Therefore, varying column density and covering factor of the absorber provide a sensible physical explanation for the observed \xray\ variability.

The rapid \xray\ variability between the \chandra\ and \xmm\ observations provides a constraint on the transverse velocity of the absorber.
The covering factor varied by $\approx50\%$ between these observations, within 32 \hbox{rest-frame} days.
Adopting a typical size of $10{R}_{\rm g}$ (${R}_{\rm g}=GM/c^{2}$ is the gravitational radius) for the \xray\ corona \citep[e.g.,][]{Dai2010,Morgan2012,Reis2013,Shemmer2014,Fabian2015}, the minimum transverse traveling distance of the absorber is thus $5{R}_{\rm g}$ (e.g., see Figure 5 of \citealt{Liu2022}).
The resulting velocity constraint is $v>{\rm 75~km/s}$.
This is a loose constraint, that does not correspond to useful constraints on the location of the absorber. 
For comparison, \cite{Liu2022} found a \hbox{two-day} variability timescale for a $6.3\times10^{9}~M_{\odot}$ SMBH mass quasar, which translates to $v\approx{0.9c}$.
More intensive \xray\ monitoring of SDSS \hbox{J$0814+5325$} might reveal shorter variability timescales.
In some cases, \hbox{partial-covering} absorption has been found to be associated with \hbox{ultra-fast} outflows (UFOs) that are typically identified from blueshifted Fe absorption lines in the \hbox{X-ray} spectra \citep[e.g.,][]{Chartas2002,Chartas2003,Hagino2016,Matzeu2016,Reeves2016,Reeves2018,Reeves2020}.
The \hbox{XMM-Newton} spectra do not show any apparent absorption lines, and thus we are unable to constrain the absorber velocity this way.

Besides the obscuration scenario, another possible mechanism that has been invoked to explain the strong \hbox{X-ray} variability in AGNs is relativistic disk reflection. 
In this scenario, changes of the corona height/size affect the strengths of the \hbox{light-bending} and \hbox{disk-reflection} effects, and a smaller corona height/size results in weaker observed \hbox{X-ray} emission as a larger fraction of the coronal emission is focused on the inner accretion disk due to the \hbox{light-bending} effects.
The low flux state spectrum is characterized by a \hbox{reflection-dominated} continuum and a broad Fe K$\alpha$ line. 
This model has had much success in explaining the spectral shapes and features of different flux states in several NLS1s \citep[e.g.,][]{Grupe2008, Fabian2012,Parker2014,Jiang2018} and even quasars \citep[e.g.,][]{Miniutti2012}.
However, these studies generally do not examine the \hbox{X-ray} variability in the context of \hbox{X-ray} emission strength relative to optical/UV emission strength (i.e., the $\alpha_{\rm OX}$--$L_{2500~\textup{\AA}}$ relation).
Thus it is unclear if varying the corona height/size alone is sufficient to account for the \hbox{X-ray} variability, and some studies \citep[e.g.,][]{Miniutti2012} have suggested that significant changes of the intrinsic coronal emission are still needed in addition to disk reflection.

For SDSS J0814+5325, the Chandra and Swift spectra have limited photon statistics, hampering detailed spectral modeling.
It is also difficult to discriminate disk reflection from \hbox{partial-covering} absorption for spectra without clear broad Fe K$\alpha$ lines \citep[e.g.,][]{Tanaka2004, Dauser2012, Waddell2019, Boller2021, Jiang2022}.
But a simple argument against the \hbox{disk-reflection} scenario is based on the $>5~{\rm keV}$ excess (Figure~\ref{fig:xray_xmm}a) and the relatively normal flux level ($\Delta\alpha_{\rm OX}=-0.13$ and $f_{\rm weak}=2.2$) of the XMM-Newton spectra.
In the \hbox{disk-reflection} scenario, a large reflection fraction is needed to account for the hard \hbox{X-ray} excess.
However, strong reflection effects should correspond to strong \hbox{light-bending} effects, and thus we would expect a \hbox{low-state} spectrum.
For example, a reflection fraction of $\approx11$ and a corona height of $\approx2~R_{\rm g}$ ($R_{\rm g}$ being the gravitational radius) were obtained if we fitted the \hbox{XMM-Newton} spectra with the \hbox{disk-reflection} model {\sc relxilllp} \citep{Dauser2014, Dauser2016}.\footnote{The inclination angle is fixed at the default value of 30\degree, and the free parameters are the photon index, reflection fraction ($R_{\rm ref}$), corona height ($h$), BH spin ($a$), ionization parameter ($\xi$), metallicity ($A_{\rm Fe}$), and normalization. The model describes well the spectra with $\chi^{2}/{\rm dof}=359.7/385$. The spectra appear reflection dominated, with \hbox{best-fit} parameters of $\Gamma=1.91^{+0.03}_{-0.03}$, $R_{\rm ref}=10.8_{-3.3}$, $h=2.0^{+0.4}~R_{\rm g}$, $a=0.89^{+0.06}_{-0.08}$, $\xi=1.2^{+0.2}_{-0.3}\times10^{3}~{\rm erg}~{\rm cm}~{\rm s}^{-1}$, and $A_{\rm Fe}=0.67^{+0.17}_{-0.67}$.}
Compared to the same model with a reflection fraction of 1 and a corona height of $10~R_{\rm g}$ (approximately a high-state spectrum), the $f_{\rm 2~keV}$ value of the \hbox{best-fit} model is 7.2 times lower.
Therefore, the \hbox{XMM-Newton} spectra ($f_{\rm weak}=2.2$) are unlikely to be explained by the \hbox{disk-reflection} model.
A similar argument may also apply to the 2010 Swift spectrum that has a relatively flat \hbox{power-law} shape ($\Gamma=1.58\pm{0.21}$; likely also requiring a large reflection fraction in the \hbox{disk-reflection} model) and is at a nominal flux level ($\Delta\alpha_{\rm OX}=-0.03$ and $f_{\rm weak}=1.2$).
Overall, after we take the varying X-ray emission strength (relative to the \hbox{$\alpha_{\rm OX}$--$L_{2500~\textup{\AA}}$ relation}) into consideration, \hbox{partial-covering} absorption appears the most plausible scenario.

\subsection{Connection to ``Intrinsically X-ray Weak'' Quasars}
\label{sec:4.2}

A small number of quasars have been proposed to be intrinsically X-ray weak, with their \xray\ coronae producing much weaker continuum emission than expected from the $\alpha_{\rm OX}$--$L_{\rm 2500~{\textup{\AA}}}$ relation; their spectra are steep ($\Gamma\gtrsim2$) power laws showing no clear evidence of \xray\ obscuration \citep[e.g.,][]{Leighly2007b,Leighly2007a,Luo2014}.
However, based on \hbox{multi-epoch} soft and hard \xray\ observations of three such candidates, \cite{Wang2022} proposed that these weak \xray\ continua may be alternatively explained with heavy obscuration, and the steep spectra are dominated by the leaked component from \hbox{partial-covering} absorption.
They also suggested that this phenomenon preferentially occurs in \hbox{super-Eddington} accreting quasars as they may drive powerful and \hbox{high-density} winds.

SDSS \hbox{J$0814+5325$} first attracted our attention from its \chandra\ observation, which showed a weak yet steep \xray\ spectrum, mimicking those of intrinsically \xray\ weak quasars.
The \hbox{follow-up} \xmm\ TOO observation a month later, however, revealed a more nominal \xray\ flux level with signatures of \hbox{partial-covering} absorption.
Therefore, SDSS \hbox{J$0814+5325$} provides another piece of evidence that apparent intrinsically \xray\ weak quasars might instead be explained with obscuration.
Owing to the accumulation of \xray\ data, an increasing number of similar quasars have been reported recently.
For example, another \hbox{reverberation-mapped} \hbox{super-Eddington} accreting quasar, SDSS \hbox{J075101.42+291419.1}, showed similar \xray\ weakness and \xray\ variability, as well as a steep spectrum in the low state \citep{Liu2019}.
Small samples of intrinsically \xray\ weak quasar candidates were also reported from X-ray surveys of luminous quasars \citep[e.g.,][]{Nardini2019,Laurenti2022}.
It is possible that these objects are uniformly explained with the simple obscuration scenario; \hbox{multi-epoch} \xray\ observations will be helpful for discerning their nature.

\section{Summary} \label{sec:5}

We report the strong and rapid \xray\ variability of the \hbox{super-Eddington} accreting quasar SDSS \hbox{J0814+5325} discovered from new \chandra\ and \xmm\ observations plus archival \swift\ observations.
The key points are summarized as follows.
\begin{enumerate}

\item
Given the light curve of the \hbox{2 keV} flux density shown in Figure~\ref{fig:ftkev}, the largest variability amplitude is a factor of $9.6^{+11.6}_{-4.6}$, between the 2011 August \swift\ observation and the 2021 \chandra\ observation.
Between the \chandra\ and \xmm\ observations in 2021, the \hbox{2 keV} flux density varied by a factor of $5.3^{+6.4}_{-2.4}$ in 32 \hbox{rest-frame} days.
See Sections~\ref{sec:3.1} and \ref{sec:3.3}.

\item
The \chandra\ spectrum is described by a simple \hbox{power-law} continuum with $\Gamma=2.48_{-0.65}^{+0.67}$.
It displays the weakest \xray\ emission with $\Delta\alpha_{\rm OX}=-0.41_{-0.13}^{+0.10}$ and $f_{\rm weak}=11.7_{-6.3}^{+9.6}$.
The \xmm\ spectra 32 days later are best described by a power law modified with \hbox{partial-covering} absorption, and the \hbox{absorption-corrected} emission is at a nominal level with $\Delta\alpha_{\rm OX,~corr}=-0.01$.
See Sections~\ref{sec:3.1} and \ref{sec:3.3}.

\item
There is no \hbox{short-term} optical spectral variability from the nearly simultaneous optical spectra. 
See Section~\ref{sec:2.4}.
The \hbox{long-term} optical/IR variability amplitudes are only mild (a maximum value of $\approx50\%$).
There is significant dust extinction ($E(B-V)\approx0.073$) inferred from the optical and \xmm\ OM data.
See Sections~\ref{sec:3.2} and~\ref{sec:3.4}.

\item
We interpret the \xray\ variability with an obscuration scenario.
The absorber is the clumpy \hbox{accretion-disk} wind having variable column density and covering factor along the line of sight.
Powerful and \hbox{high-density} winds are expected in \hbox{super-Eddington} accreting quasars.
This scenario may uniformly explain SDSS \hbox{J$0814+5325$} and other \hbox{super-Eddington} accreting quasars displaying similar \xray\ and multiwavelength properties.
See Section~\ref{sec:4}.

\end{enumerate}

For SDSS \hbox{J$0814+5325$}, the \hbox{high-quality} \xmm\ \hbox{follow-up} observation provided key information (i.e., an intrinsic \hbox{nominal-level} continuum affected by \hbox{partial-covering} absorption) for understanding the unusually weak yet steep \chandra\ spectrum.
Thus \hbox{multi-epoch} \hbox{high-quality} \xray\ observations appear important to reveal the nature of other quasars displaying similar \xray\ weakness and/or \xray\ variability. 

\vspace{5mm}

We acknowledge \xmm\ for obtaining the TOO observation.
J.H. and B.L. acknowledge financial support from the National Natural Science Foundation of China grant 11991053, China Manned Space Project grants NO. \hbox{CMS-CSST-2021-A05} and NO. \hbox{CMS-CSST-2021-A06}.
W.N.B. acknowledges support from the Chandra X-ray Center grant \hbox{GO2-23083X} and Penn State ACIS Instrument Team Contract SV4-74018 (issued by the Chandra \xray\ Center, which is operated by the Smithsonian Astrophysical Observatory for and on behalf of NASA under contract NAS8-03060).
J.M.W. acknowledges financial support from the National Natural Science Foundation of China grants NSFC-11991050, -11991054, and -11833008.
The Chandra ACIS Team Guaranteed Time Observations (GTO) utilized were selected by the ACIS Instrument Principal Investigator, Gordon P. Garmire, currently of the Huntingdon Institute for \xray\ Astronomy, LLC, which is under contract to the Smithsonian Astrophysical Observatory via Contract SV2-82024.
We acknowledge the support of the staff of the Lijiang 2.4 m telescope (LJT). Funding for the telescope has been provided by Chinese Academy of Sciences and the People’s Government of Yunnan Province.

\bibliographystyle{aasjournal}
\bibliography{ms.bib}

\appendix
The eight targets (including SDSS \hbox{J0814+5325}) in our \chandra\ snapshot program are optically bright SDSS DR7 quasars with $m_{i}<17.4$, and they are chosen from a parent sample of 24 \hbox{super-Eddington} accreting AGNs that have been monitored by \hbox{reverberation-mapping} campaigns \citep[e.g.,][]{Du2015,Du2016,Du2018}.
These eight quasars had limited numbers of previous \xray\ observations with short exposures (mostly from \swift), and the main aim of the snapshot program is to search for strong \xray\ variability among \hbox{super-Eddington} accreting AGNs.
We adopted the same procedure as described in Section~\ref{sec:2.1} to reduce the \chandra\ data of the other seven targets.
All seven quasars were significantly detected by \chandra.
We used a simple \hbox{power-law} model modified by Galactic absorption to fit the \hbox{$0.3\textrm{--}8~{\rm keV}$} spectra.
The \hbox{rest-frame} \hbox{2 keV} flux densities were then calculated from the \hbox{best-fit} models.
We collected their $L_{2500~\textup{\AA}}$ values from \cite{Liu2021} or the SDSS DR7 quasar catalog \citep{Shen2011}, and we then computed their $\alpha_{\rm OX}$ and $\Delta\alpha_{\rm OX}$ values.
The results for the other seven quasars are shown in Table~\ref{tbl:gto}.
Among these objects, \hbox{SDSS J1023+5233} showed nominal level of \xray\ emission in the \chandra\ observation, but its \hbox{2 keV} flux density from its archival \swift\ observation is $\approx8.0$ times lower.
\hbox{SDSS J0759+3200} showed weak \xray\ emission in the \chandra\ observation, and its 2 keV flux density is $\approx4.3$ times lower than that of its archival Swift observation.
We present only the results for \hbox{SDSS J0814+5325} in this study as it has   \xmm\ and Lijiang \hbox{follow-up} observations.
Statistical investigations of strong \xray\ variability among the sample of \hbox{super-Eddington} accreting AGNs are to be carried out in future studies.

\begin{deluxetable*}{lcccccccrr}
\tabletypesize{\scriptsize}
\tablecolumns{10}
\tablecaption{\xray\ and Optical/UV Properties of the Other Seven Quasars in the \chandra\ Snapshot Program}
\tablehead{
\colhead{Object Name}  &
\colhead{Redshift}  &
\colhead{${L_{2500~\textup{\AA}}}$} &
\colhead{Observation} &
\colhead{Exposure} &
\colhead{Spectral} &
\colhead{$\Gamma$} &
\colhead{${f_{2~{\rm keV}}}$} &
\colhead{$\alpha_{\rm OX}$} &
\colhead{$\Delta\alpha_{\rm OX}$}  \\
\colhead{(J2000)}   &
\colhead{ }   &
\colhead{($10^{29}~{\rm erg}~{\rm cm}^{-2}~{\rm s}^{-1}$)}   &
\colhead{ ID } &
\colhead{Time (ks)} &
\colhead{Counts} &
\colhead{ }   &
\colhead{($10^{-31}~{\rm erg}~{\rm cm}^{-2}~{\rm s}^{-1}$)}   &
\colhead{ }   &
\colhead{ } \\
\colhead{(1)}   &
\colhead{(2)}   &
\colhead{(3)}   &
\colhead{(4)}   &
\colhead{(5)}   &
\colhead{(6)}   &
\colhead{(7)}   &
\colhead{(8)}   &
\colhead{(9)}   &
\colhead{(10)} 
}
\startdata
075051.72+245409.3 & 0.400 & 27.27  & 24723 &   3.4 & 122  & $ 2.22^{+0.19}_{-0.19} $ & $6.05^{+1.51}_{-1.31}$ & $-1.54^{+0.04}_{-0.04}$ &  -0.01 \\
075949.54+320023.8 & 0.188 & 3.16  & 24724 &   1.9 & 24 & $ 1.72^{+0.45}_{-0.45} $ & $1.67^{+1.02}_{-0.74}$ & $-1.66^{+0.08}_{-0.10} $ & -0.25 \\
080101.41+184840.7 & 0.140 & 3.39  & 24725 &   1.4 & 181 & $ 2.26^{+0.16}_{-0.16} $ & $15.90^{+2.76}_{-2.51}$ &  $-1.40^{+0.03}_{-0.03}$ &  0.01  \\
081441.91+212918.5 & 0.163 & 3.09  & 24726 &   1.4 & 77  & $ 1.78^{+0.23}_{-0.22} $ & $6.74^{+1.96}_{-1.66}$ & $-1.47^{+0.04}_{-0.05}$ & -0.07 \\
085946.35+274534.8 & 0.244 & 2.95  & 24728 &   1.9 & 67  & $ 2.03^{+0.26}_{-0.26} $ & $4.92^{+1.62}_{-1.36}$ & $-1.38^{+0.05}_{-0.05}$ & 0.02 \\
093922.89+370943.9 & 0.186 & 2.24  & 24729 &   2.0 & 38  & $ 1.89^{+0.32}_{-0.31} $ & $2.37^{+0.98}_{-0.79}$ & $-1.55^{+0.06}_{-0.07}$ & -0.16 \\
102339.64+523349.6 & 0.136 & 1.05  & 24730 &   1.9 & 155  & $ 1.91^{+0.16}_{-0.16} $ & $9.99^{+1.90}_{-1.71}$ & $-1.29^{+0.03}_{-0.03}$ &  0.05 \\
\enddata
\tablecomments{
Column (1): name of the object, in order of increasing right ascension.
Column (2): redshift.
Column (3): luminosity at \hbox{rest-frame} \hbox{2500~\AA} from either \cite{Liu2021} or \cite{Shen2011}.
Column (4): \chandra\ observation ID.
Column (5): cleaned exposure time.
Column (6): net spectral counts.
Column (7): \xray\ photon index from the \hbox{best-fit} simple \hbox{power-law} model.
Column (8): flux density at \hbox{rest-frame} \hbox{2~keV}.
Column (9): observed $\alpha_{\rm OX}$ value.
Column (10): difference between the observed and expected $\alpha_{\rm OX}$ derived from the \cite{Steffen2006} $\alpha_{\rm OX}\textrm{--}L_{\rm 2500~{\textup{\AA}}}$ relation.
}
\label{tbl:gto}
\end{deluxetable*}

\end{document}